\pdfoutput=1
\documentclass[aps,prc,onecolumn]{revtex4}
\usepackage{epsf,epsfig,amssymb,amsmath}

\usepackage{color}

\newcommand\SKIP[1]{}
\newcommand{\be}{\begin{equation}}
\newcommand{\ee}{\end{equation}}
\newcommand{\bea}{\begin{eqnarray}}
\newcommand{\eea}{\end{eqnarray}}
\newcommand{\mybibitem}{\bibitem}

\newcommand{\ch}{{\rm ch}}
\newcommand{\sh}{{\rm sh}}

\newcommand{\gton}{\mathrel{\lower.9ex \hbox{$\stackrel{\displaystyle 
>}{\sim}$}}} 
\newcommand{\lton}{\mathrel{\lower.9ex \hbox{$\stackrel{\displaystyle 
<}{\sim}$}}}

\newcommand{\vx}{{\bf x}}

\newcommand{\vn}{{\bf n}}
\newcommand{\vp}{{\bf p}}
\newcommand{\vq}{{\bf q}}
\renewcommand{\vr}{{\bf r}}
\newcommand{\vv}{{\bf v}}
\newcommand{\vzero}{{\bf 0}}

\newcommand{\ttau}{{\tilde \tau}}

\newcommand{\myell}{{\ell}}
\newcommand{\feq}{f^{\rm eq}}

\begin{document}

\title{Self-consistent conversion of a viscous fluid to particles}

\author{Denes Molnar}
\author{Zack Wolff}
\affiliation{Department of Physics and Astronomy, Purdue University, West Lafayette, IN 47907}

\date{\today}

\begin{abstract}
Comparison of hydrodynamic and ``hybrid'' hydrodynamics+transport 
calculations to heavy-ion data 
inevitably requires the
conversion of the fluid to particles. For dissipative fluids the conversion
is ambiguous without additional theory input complementing hydrodynamics.
We obtain self-consistent shear viscous phase space corrections 
from linearized Boltzmann transport theory for a gas of hadrons.
These corrections depend on the particle
species, and incorporating them in Cooper-Frye freezeout 
affects identified particle observables.
For example, with additive quark model cross sections,
proton elliptic flow is larger than pion elliptic flow at moderately
high $p_T$ in $Au+Au$ collisions at RHIC.
This is in contrast to Cooper-Frye freezeout with the commonly used 
``democratic Grad'' ansatz that assumes no species dependence. 
Various analytic and numerical results are also presented
for massless and massive two-component mixtures to better elucidate 
how species dependence arises.
For convenient inclusion in 
pure hydrodynamic and hybrid calculations, Appendix G contains
self-consistent viscous corrections for each species both in tabulated and
parameterized form.
\end{abstract}

\maketitle

\section{Introduction}

The most common dynamical 
framework to interpret data from ultrarelativistic heavy-ion ($A+A$) reactions
is relativistic hydrodynamics\cite{hydro}. Application of hydrodynamics
necessitates the conversion of the fluid to particles, which are then either
evolved further in a hadronic transport model or assumed to free stream
to the detectors. The usual approach to such ``particlization'' 
\cite{HuovinenPetersen} is to do the conversion
on a constant temperature or energy density hypersurface in spacetime 
via the Cooper-Frye formula\cite{CooperFrye}.
While unambiguous for fluids in perfect local thermal
equilibrium, i.e., {\em ideal} fluids, for {\em dissipative} fluids an 
infinite class of phase space densities can reproduce the same hydrodynamic
fields.
This is further exacerbated for mixtures where one can postulate
phase space corrections for each particle species almost independently.

In practice these ambiguities are commonly ignored, even
in state-of-the-art ``hybrid'' hydro+transport calculations\cite{VISHNU}. 
For example,
 shear viscous corrections are simply assumed to follow quadratic momentum 
dependence with a common coefficient for all species, a procedure one of us 
termed ``democratic Grad'' ansatz\cite{dyngrad}. This, however, ignores
the very microscopic dynamics that keeps the hadron gas near local 
equilibrium. We apply here instead a self-consistent approach
that obtains shear viscous corrections from linearized kinetic theory for
a gas of hadrons. This extends earlier studies that considered 
massless quarks and gluons\cite{Dusling_deltaf}, or hadronic mixture with
two species only\cite{dyngrad}.

Recently there has been a lot of interest in bulk viscous corrections 
\cite{Monnai:2009ad,Dusling:2011fd,Noronha-Hostler:2013gga}.
While this work focuses on phase space corrections due to shear only, 
the technique used here could be extended to the bulk viscous
case in a straightforward manner. Shear corrections also affect
photon and dileption emission from the quark-gluon plasma 
in heavy-ion collisions\cite{Schenke:2006yp}.

For simplicity we consider phase space corrections with power-law
momentum dependence, most prominently the quadratic Grad form, 
so that the corrections can be simply represented by numbers 
(instead of numerically determined functions). 
This will be remedied in a future publication.
General aspects of the approach are presented in Section~\ref{Sc:framework}, 
followed by analytic and numerical results for
massless and massive two-component mixtures in 
Sections~\ref{Sc:massless} and~\ref{Sc:massive_2comp},
and numerical results on 
the particle species dependence of differential 
elliptic flow $v_2(p_T) \equiv
\langle \cos 2\phi\rangle_{p_T}$
for a multicomponent hadronic gas
in Section~\ref{Sc:multi}.  
The approach is also verified against fully nonlinear kinetic theory
in Section~\ref{Sc:BTE_2comp}.
Technical details are deferred to Appendices A-F. We only highlight
here Appendix~\ref{App:tables}, which contains tables and parameterizations of
self-consistent species-dependent
correction factors to the commonly used 
``democratic'' Cooper-Frye freezeout.
These facilitate implementation of our results in hydrodynamic and hybrid
calculations.

\section{Viscous phase space corrections from linearized transport}
\label{Sc:framework}

\subsection{Democratic Grad ansatz}
\label{Sc:dem_grad}

The principle challenge in converting a fluid to particles is that one needs 
to obtain phase space densities
\be
f_i(x,\vp) \equiv \frac{dN_i(\vr,\vp,t)}{d^3r\, d^3p}
\ee
for each of the particle species $i$ solely from hydrodynamics fields,
namely
the energy-momentum tensor $T^{\mu\nu}$ and any conserved charge
currents $N_c^\mu$ (in heavy-ion physics applications, 
typically the baryon charge).
The conversion is envisioned in spacetime regions where
the hydrodynamic and particle descriptions are to good approximation 
equivalent, so we only 
switch 'language' but the state of the system is unchanged\cite{FOlayer}.
The particles are usually modeled as a gas, in which case
one has to invert%
\footnote{
High-energy physics units $\hbar = c = k_B = 1$ and the metric 
with $(+,-,-,-)$ signature are used throughout,
with Einstein conventions in all Lorentz tensor expressions,
and Minkowski scalar products abbreviated as $(ab)\equiv a_\mu b^\mu$.
Sums over particle species, on the other hand, are always written explicitly.
}
\be
T^{\mu\nu}(x) \equiv \sum\limits_i \int\limits \frac{d^3 p}{E} p^\mu p^\nu f_i(x,\vp)
\label{Tmunu_def}
\ee 
and 
\be
N_c^\mu(x) \equiv \sum\limits_i q_{c,i} \int\limits  
                   \frac{d^3 p}{E} p^\mu f_i(x,\vp) \ ,
\label{Nmu_def}
\ee 
where $q_{c,i}$ is the charge of type $c$ carried by a particle of species $i$.

For {\em non}dissipative fluids, which by definition are 
in local equilibrium everywhere in space at all times,
the conversion is straightforward
because in local thermal and chemical equilibrium 
particle distributions are%
\footnote{
Throughout this paper 
Boltzmann statistics is assumed but generalization
to the Bose/Fermi case is straightforward.
}
\be
f_i(x,\vp) \equiv
 \feq_i(x, \vp) = \frac{g_i}{(2\pi)^3} \exp\!\left[
                              \frac{\mu_i(x) - p_\alpha u^\alpha(x)}{T(x)}
                                     \right]   \quad ,  \qquad
\mu_i \equiv \sum\limits_c q_{c,i} \mu_c(x) \ ,
\label{feq}
\ee
where $g_i$ is the number of internal degrees of freedom for species 
$i$. The combination $p_\alpha u^\alpha$ is
the energy of the particle in the local rest (LR) frame of the fluid
($u^\mu_{LR} = (1,\vzero)$). 
The local temperature $T$, chemical potentials $\{\mu_c\}$, 
and four-velocity $u^\mu$
of fluid flow are uniquely determined through the ideal hydrodynamic
relations
\be
T_{\rm id}^{\mu\nu}(x) = [e(x) + p(x)] u^\mu(x) u^\nu(x) - p(x) g^{\mu\nu}
\quad , \qquad
N_{c,\rm id}^\mu(x) = n_c(x) u^\mu(x) \ ,
\label{id_hydro_T_N}
\ee
with rest frame energy density $e(T,\{\mu_c\})$, pressure $p(T,\{\mu_c\})$, and charge density $n_c(T,\{\mu_c\})$
given by the equation of state (these can be inverted for
$T$ and $\{\mu_c\}$). 
For consistency,
at the point of conversion
the equation of state used in fluid dynamics must of course
correspond to a gas of particles.

If the fluid is dissipative, then it is not strictly in 
local thermal and chemical equilibrium, and 
phase space densities therefore acquire dissipative
corrections
\be
f_i(x,\vp) = \feq_i(x,\vp) + \delta f_i(x, \vp)  
\equiv \feq_i(x,\vp) [1 + \phi_i(x,\vp)]\ .
\label{def_deltaf}
\ee
The ideal hydrodynamic forms (\ref{id_hydro_T_N}) no longer hold because
the energy-momentum tensor and charge currents acquire nonideal corrections
\be
T^{\mu\nu} = T_{id}^{\mu\nu} + \delta T^{\mu\nu} \quad \ , \qquad
N_c^\mu = N_{c,\rm id}^\mu + \delta N_c^\mu \quad \quad\qquad
(u_\mu \delta T^{\mu\nu} u_\nu = 0, \ u_\mu \delta N_c^\mu = 0)\ ,
\ee
where $\delta T^{\mu\nu}$ is customarily decomposed further 
into a shear stress tensor $\pi^{\mu\nu}$ and bulk pressure $\Pi$:
\be
\delta T^{\mu\nu} = \pi^{\mu\nu} + \Pi (u^\mu u^\nu - g^{\mu\nu}) \quad ,
\qquad
\pi^\mu_\mu \equiv 0 \ ,
\ee
if one uses Landau convention for fluid flow definition 
(so $u_\mu \delta T^{\mu\nu} \equiv 0$). 
On the other hand,
(\ref{Tmunu_def}) and (\ref{Nmu_def}) remain valid and can be recast as
\be
\delta T^{\mu\nu}(x) = \sum\limits_i \int\limits \frac{d^3 p}{E} p^\mu p^\nu \delta f_i(x,\vp) \quad \ , \qquad
\delta N_c^\mu(x) = \sum\limits_i q_{c,i} \int\limits
                   \frac{d^3 p}{E} p^\mu \delta f_i(x,\vp) \ .
\label{df_constraints}
\ee 
Without additional information about the functional form of the $\delta f_i$,
this finite set of conditions can be satisfied with infinitely 
many different $\delta f_i$ (or equivalently, $\phi_i$), 
even if there is only a single particle species.

Often the only dissipative correction considered is shear stress.
A common prescription that satisfies the constraint (\ref{df_constraints})
from shear
is the ``democratic Grad'' ansatz\cite{dyngrad}, which assumes 
phase space corrections with quadratic momentum dependence
\be
\phi_i^{\rm dem}(x,\vp) = \frac{\pi^{\mu\nu}(x)p_\mu p_\nu}{2[e(x)+p(x)]T^2(x)}
\ .
\label{phi_dem}
\ee
Note, the coefficient in this quadratic form is 
the same for all particle species. 
The reason this ansatz
works is that for each species it 
gives a partial shear stress that is proportional to the partial enthalpy:
\be
\pi_i^{\mu\nu} \equiv \int \frac{d^3 p}{E} p^\mu p^\nu \delta f_i^{\rm dem} 
= \frac{e_i + p_i}{e + p} \pi^{\mu\nu}
\qquad\Rightarrow \qquad \sum\limits_i \pi_i^{\mu\nu} = \pi^{\mu\nu} \ .
\ee
However, this simple choice
ignores the very microscopic dynamics that keeps the gas near
local equilibrium. 
In particular, one expects species that interact more frequently
to be better equilibrated than those that scatter less often.

\subsection{Covariant transport theory}
\label{Sc:cov_trans}

In contrast, a 
self-consistent set of dissipative corrections can be obtained from 
linearized covariant transport theory.
Consider on-shell covariant transport theory for a multicomponent system with
$2\to 2$ interactions. For each species $i$ the evolution of
the phase space density is given by the nonlinear Boltzmann transport
equation
\be
p^\mu \partial_\mu f_{i}(x,\vp) = S_i(x,\vp) + \sum\limits_{jk\myell} 
C^{ij\to k\myell}[f_i,f_j,f_k,f_\myell](x, \vp) \ ,
\label{BTE}
\ee
where the source term $S_i$ encodes the initial conditions,
and the collision terms are%
\footnote{
In (\ref{Cijkl}) outgoing momenta $p_3$ and $p_4$ 
are understood to be integrated over full, unrestricted phasespace.
This double-counts the rate for identical particles ($k = \ell$) compared
to nonidentical particles, 
however, that is compensated by double-counting in the sum for 
$k \ne \ell$. See also (\ref{int34_W}).
}
\be
C^{ij\to k\myell}[f_i,f_j,f_k,f_\myell](x,\vp_1) 
\equiv \int\limits_2 \!\!\!\!\int\limits_3 \!\!\!\!\int\limits_4
\left(\frac{g_i g_j}{g_k g_\myell} f_{3k} f_{4\myell} - f_{1i} f_{2j}\right)
\, \bar W_{12\to 34}^{ij\to k\myell}  \, \delta^4(12 - 34)
\label{Cijkl}
\ee
with shorthands
$\int\limits_a \equiv \int d^3p_a / (2 E_a)$, 
$f_{ai} \equiv f_i(x,\vp_a)$, and
$\delta^4(ab - cd) \equiv \delta^4(p_a + p_b - p_c - p_d)$.
The transition probability $\bar W_{12\to 34}^{ij\to k\myell}$ 
for the process $i + j \to k + \ell$ with momenta 
$p_1 + p_2 \to p_3 + p_4$
is invariant under interchange of incoming or outgoing particles,
\be
 \bar W_{12\to 34}^{ij\to k\myell} \equiv \bar W_{21\to 34}^{ji \to k\myell}
 \equiv \bar W_{12\to 43}^{ij \to \myell k} 
 \equiv \bar W_{21\to 43}^{ji \to \myell k} \ ,
\label{W_symmetry}
\ee
satisfies
detailed balance
\be
 \bar W_{34\to 12}^{k\myell \to ij} 
      \equiv \frac{g_i g_j}{g_k g_\myell} \bar W_{12\to 34}^{ij\to k\myell} \ ,
\label{W_balance}
\ee
and is given by the corresponding unpolarized scattering matrix element
or differential cross section as 
\be
\bar W_{12\to 34}^{ij\to k\myell} 
 = \frac{1}{16\pi^2} |\overline {{\cal M}_{12\to 34}^{ij\to k\myell}}|^2
 \equiv \frac{4}{\pi} s p_{cm}^2 \frac{d\sigma_{12\to 34}^{ij\to k\myell}}{dt}
\equiv 4s \frac{p_{cm}}{p'_{cm}} 
      \frac{d\sigma_{12\to 34}^{ij\to k\myell}}{d\Omega_{cm}} \ .
\label{W_with_sigma}
\ee
Here $s \equiv (p_1 + p_2)^2$ and $t \equiv (p_1 - p_3)^2$ 
are standard Mandelstam variables, while 
\be
p_{cm} \equiv \frac{\sqrt{(p_1 p_2)^2 - m_i^2 m_j^2}}{\sqrt{s}} \quad \ ,
\qquad p'_{cm} \equiv \frac{\sqrt{(p_3 p_4)^2 - m_k^2 m_\myell^2}}{\sqrt{s}}
\ee
are the magnitudes of incoming and outgoing particle momenta 
in the center of mass frame of the microscopic two-body collision.
The degeneracy factors $g$ of the species 
appear explicitly in (\ref{W_balance}) 
because unpolarized matrix
elements are summed over internal degrees of freedom 
(spin, polarization, color) of outgoing particles, whereas 
{\em averaged} over those of incoming particles. These factors also
appear in (\ref{Cijkl}) because distribution
functions here are assumed to depend only on momentum and position 
but not on internal degrees of freedom, and thus the distribution of each
species is summed over internal degrees of freedom 
(cf. the local equilibrium form (\ref{feq})).

\subsection{Self-consistent viscous 
corrections from linearized covariant transport}
\label{Sc:lin_trans}

For small departures from local equilibrium one can split each phase
space density into a local equilibrium part and a dissipative correction
as in (\ref{def_deltaf}),
and linearize (\ref{BTE}) in $\delta f$:
\be
p^\mu \partial_\mu \feq_i + p^\mu \partial_\mu \delta f_{i}
= \sum\limits_{jk\myell} 
    \left\{C^{ij\to k\myell}[\delta f_i, \feq_j,\delta f_k, \feq_\myell]
         + C^{ij\to k\myell}[\feq_i, \delta f_j,\feq_k, \delta f_\myell]
\right\}
\label{linBTE}
\ee
(with the source term dropped and spacetime and momentum arguments suppressed).
 The 
solutions to this coupled set of equations, of course, depend on both
the matrix elements and initial conditions. However, typical
systems quickly 
relax on microscopic scattering timescales to a solution dictated
by gradients of the equilibrium distribution on the left hand side of 
(\ref{linBTE}). The asymptotic solution, for given gradients, is then 
uniquely determined by the interactions in the system 
(to see this relaxation worked out explicitly,
check Ref.~\cite{Denicol:2012cn}).
In this so-called Navier-Stokes regime, one can neglect the time derivative of 
$\delta f_i$, and if gradients of $\feq_i$ are small, 
one can also ignore%
\footnote{
$\delta f_i$ is to leading order proportional to the gradients of $\feq_i$,
and if those are small due to a large length scale $L$ in the problem
$\nabla_\mu \feq_i \sim 1/L$,
then $\nabla_\mu \delta f_i \sim 1/L^2$ is suppressed compared to 
$\nabla_\mu f_i$.
}
the spatial derivatives of $\delta f_i$. At each spacetime
point $x$ one then has a linear integral equation to solve. 
This is also the starting point of the standard calculation of 
transport 
coefficients
in kinetic theory\cite{trcoeffs}.
For example, the shear viscosity $\eta_s$ and bulk viscosity 
$\zeta$ are defined in the Navier-Stokes limit through
\be
\delta T^{\mu\nu}_{NS} \equiv 
\eta_s \sigma^{\mu\nu} + \zeta \Delta^{\mu\nu} 
(\partial u)  \quad , \qquad 
\sigma^{\mu\nu} \equiv \nabla^\mu u^\nu + \nabla^\nu u^\mu 
                 - \frac{2}{3}\Delta^{\mu\nu} (\partial u) \ ,
\label{trcoeff_def}
\ee 
where $\Delta^{\mu\nu} \equiv g^{\mu\nu} - u^\mu u^\nu$ is a convenient 
projector to isolate spatial derivatives 
$\nabla^\mu = \Delta^{\mu\nu} \partial_\nu$
in the local rest (LR) frame.

The derivative on the LHS of (\ref{linBTE}) can be written as
\bea
(p\partial) \feq_i
  = &\feq_i& \left\{p_\alpha \left[
                   \nabla^\alpha \frac{\mu_i}{T} 
                    - (p u) \nabla^\alpha \frac{1}{T}
                   \right]
                   + (p u) (u\partial) \frac{\mu_i}{T}
                   - (p u)^2 (u\partial) \frac{1}{T} \right.
\nonumber \\
&&\qquad \left. - \, \frac{p_\alpha p_\beta}{2T}
               \left[
                 \left(\nabla^\alpha u^\beta + \nabla^\beta u^\alpha
                       - \frac{2}{3}\Delta^{\alpha\beta}(\partial u)\right)
                       + \frac{2}{3}\Delta^{\alpha\beta}(\partial u) 
                 \right]
               - \frac{(p u)}{T} p_\alpha (u\partial) u^\alpha
          \right\} \ .
\label{linBTE_source}
\eea
To isolate the response to shear, 
take uniform temperature and chemical potentials $T=const$, $\mu_c = const$,
with $\sigma^{\mu\nu} \ne 0$
but $(\partial u) = 0$. Only terms on the second line remain;
the ones in the square bracket contribute to $\delta T^{\mu\nu}$,
whereas the last term with temporal derivative $(u\partial)$ 
can be dropped as long as gradients are weak%
\footnote{
Time derivatives of hydrodynamic quantities can be replaced with spatial ones
using the energy-momentum and charge conservation laws 
$\partial_\mu T^{\mu\nu}(x) = 0$, 
$\partial_\mu N_c^\mu(x) = 0$. For example,
$$
(u\partial) u^\nu 
   = \frac{1}{e+p} [\nabla^\nu p 
                    - \nabla_\mu \delta T^{\mu\nu}
                    + \delta T^{\mu\nu} (u\partial) u_\mu]
\ , \quad
(u \partial) n_c = - n_c (\nabla u) - \nabla_\mu \delta N_c^\mu 
               + \delta N_c^\mu (u\partial) u_\mu \ ,
$$
and note that $(\partial u) \equiv (\nabla u)$. 
In shear viscosity calculations,
 pressure, energy density, and charge densities are uniform by assumption,
so derivatives of those vanish as well as derivatives of $T$ and $\mu_c$. 
What remains are
first derivatives of dissipative corrections, and dissipative corrections
times first derivatives of ideal hydrodynamic fields.
In the Navier-Stokes
regime these are of the same order and 
correspond to second derivatives of the ideal fields.
}.
With symmetric, traceless, purely spatial (in LR), and dimensionless tensors 
\be
P^{\mu\nu} \equiv \frac{1}{T^2}\left[
        \Delta^\mu_\alpha \Delta^\nu_\beta p^\alpha p^\beta - 
          \frac{1}{3}\Delta^{\mu\nu} (\Delta_{\alpha\beta} p^\alpha p^\beta)
        \right]
\ , \quad
X^{\mu\nu} \equiv \frac{\sigma^{\mu\nu}}{T} 
 = \frac{\pi^{\mu\nu}_{NS}}{\eta_s T}\ ,
\label{def_P_X}
\ee
we then  have
\be
(p\partial)\feq_i = - \frac{T^2}{2} \feq_i P^{\mu\nu}(p) X_{\mu\nu}(x) \ .
\label{source_shear}
\ee

The RHS of (\ref{linBTE}) simplifies upon the realization 
(see Appendix~\ref{App:solveChi} and Refs.~\cite{deGroot,AMYtrcoeffs}) that
\be
\phi_i(x,\vp)
   = \chi_i(|\tilde\vp|) P^{\mu\nu} X_{\mu\nu}
\quad \quad {\rm with} \quad
\left.\frac{1}{T}\Delta^{\mu\nu} p_\nu\right|_{LR} \equiv (0, \tilde\vp) \ ,
\label{chi_def} 
\ee
where $\tilde\vp$ is the LR frame three-momentum normalized by temperature.
This means that $\delta f_i$ are solely determined by real, dimensionless 
scalar functions $\chi_i$ of the rescaled momentum.
Substituting (\ref{chi_def}) and (\ref{source_shear}) 
into (\ref{linBTE}) yields, with the help of
\be
\frac{g_i g_j}{g_k g_\myell} \feq_{3k} \feq_{4\ell} \delta^4(12-34) 
\equiv \feq_{1i} \feq_{2j} \delta^4(12-34) \ ,
\label{feq_convert}
\ee
the integral equation
\be
-\frac{1}{2} P_1^{\mu\nu} \feq_{1i} 
 = \frac{1}{T^2} \sum_{jk\myell} 
            \int\limits_2\!\!\!\!\int\limits_3\!\!\!\!\int\limits_4
         \feq_{1i} \feq_{2j} \, \bar W_{12\to 34}^{ij\to k\myell}\,
         \delta^4(12-34)\, (  \chi_{3k} P_3^{\mu\nu}
                          + \chi_{4\myell} P_4^{\mu\nu}
                          - \chi_{1i} P_1^{\mu\nu}
                          - \chi_{2j} P_2^{\mu\nu}) \ ,
\ee
which after contraction with $P_{1,\mu\nu}$ reads
\be
-\frac{1}{2} P_1 \cdot P_1 \feq_{1i} 
 = \frac{1}{T^2} \sum_{jk\myell} 
            \int\limits_2\!\!\!\!\int\limits_3\!\!\!\!\int\limits_4
         \feq_{1i} \feq_{2j} \, \bar W_{12\to 34}^{ij\to k\myell}\,
         \delta^4(12-34)\, (  \chi_{3k} P_3 \cdot P_1 
                          + \chi_{4\myell} P_4 \cdot P_1
                          - \chi_{1i} P_1 \cdot P_1
                          - \chi_{2j} P_2 \cdot P_1)
\label{chi_eq}
\ee
if one introduces the notation
\be
\chi_{ai} \equiv \chi_i(|\tilde \vp_a|)
\quad , \qquad
P_a\cdot P_b \equiv P_a^{\mu\nu} P_{b,\mu\nu} 
  = (\tilde \vp_a \tilde\vp_b)^2
                       - \frac{1}{3}|\tilde \vp_a|^2 |\tilde \vp_b|^2 \ .
\ee
It is straightforward to show with the help of
(\ref{W_symmetry}), (\ref{W_balance}) and (\ref{feq_convert}) 
that (\ref{chi_eq}) is equivalent to the extremization of the functional
\bea
Q[\chi] &=& \frac{1}{2T^2} \sum\limits_i \int\limits_1 
                            P_1 \cdot P_1 \feq_{1i} \chi_{1i}
\nonumber\\
         && +\ \frac{1}{2T^4} \sum\limits_{ijk\myell}
                        \int\limits_1\!\!\!\!\int\limits_2\!\!\!\!
                        \int\limits_3\!\!\!\!\int\limits_4
           \feq_{1i} \feq_{2j} \, \bar W_{12\to 34}^{ij\to k\myell}\,
           \delta^4(12-34)\,
          (  \chi_{3k} P_3 \cdot P_1 
                          + \chi_{4\myell} P_4 \cdot P_1
                          - \chi_{1i} P_1 \cdot P_1
                          - \chi_{2j} P_2 \cdot P_1) \chi_{1i}
\nonumber \\
&\equiv& \sum_i B_i 
          + \sum_{ijk\myell} (Q_{31}^{ij\to k\myell} + Q_{41}^{ij\to k\myell}
                           - Q_{11}^{ij\to k\myell} - Q_{21}^{ij\to k\myell}) 
     \ ,
\label{Qdef}
\eea
i.e., (\ref{chi_eq}) is reproduced by the usual variational procedure imposing
$\delta Q[\chi] = 0 + {\cal O}(\delta\chi^2)$. This allows one to
estimate $\chi_i$ variationally using a finite basis $\{\Psi_{i,n}\}$ as
\be
\chi_i(|\tilde\vp|)
    = \sum\limits_n c_{i,n} \Psi_{i,n}(|\tilde\vp|)
\label{chi_expand}
\ee
and finding optimal coefficients $\{c_{i,n}\}$ that maximize $Q$
(one can in principle use different $\Psi_n$ for different species).
If the basis is complete, the limit $n\to \infty$ reproduces the 
exact solution. Numerical evaluation of $Q$ is discussed in 
Appendix~\ref{App:Qintegrals}.

The extremal value of $Q$ is directly related to the shear viscosity.
Comparison of (\ref{trcoeff_def}) to (\ref{df_constraints}) 
with (\ref{chi_def}) gives
\be
\eta_s = \frac{T_{LR}^{xz}}{\sigma_{LR}^{xz}} 
     = \frac{2}{15T^3} \sum\limits_i \int \frac{d^3 p }{E} p^4 \feq_i \chi_i 
     = \frac{4T^3}{5} \sum\limits_i B_i
\ee
with $B_i$ from (\ref{Qdef}).
On the other hand, from (\ref{chi_eq}) it follows that for the exact solution
\be
-\sum_i B_i 
  = 2 \sum_{ijk\myell} (Q_{31}^{ij\to k\myell} + Q_{41}^{ij\to k\myell}
                        - Q_{11}^{ij\to k\myell} - Q_{21}^{ij\to k\myell}) \ ,
\ee
i.e., the maximum of $Q$ is $Q_{max} = \sum\limits_i B_i /2$.
Thus, the shear viscosity is 
\be
\eta_s = \frac{8}{5} Q_{max} T^3 \ .
\label{eta_from_Q}
\ee

From (\ref{chi_def}), (\ref{chi_expand}) and (\ref{def_P_X}) one concludes that 
the democratic Grad ansatz (\ref{phi_dem}) 
corresponds to a single momentum
independent (constant) basis function with coefficient $c_i = 1$, i.e.,
\be
\chi_i^{\rm dem} = c_i \frac{\eta_s T}{2(e+p)} = \frac{\eta_s}{2s}
\label{psi_with_c}
\ee
for all species, where in the last step 
the thermodynamic identity 
$Ts = e+p - \sum\limits_{c} \mu_c n_c$ was employed
with vanishing chemical potentials 
appropriate for the midrapidity region in
heavy-ion collisions at RHIC and LHC 
energies. Except for the factor of $1/2$, the common 
$\chi_i^{\rm dem}$ value is then
just the shear viscosity to entropy ratio.
In the following we study the species dependence of $\chi_i$ 
from microscopic dynamics.

\section{Massless two-component system}
\label{Sc:massless}

Consider the so-called Grad approximation, 
in which  
\be
\chi_i(|\tilde\vp|) = \chi_i^{Grad} = const \ ,
\label{Grad}
\ee 
i.e., phase space corrections 
$\phi_i$ are quadratic in momentum.
For massless particles
with energy-independent, isotropic cross sections, the terms in
$Q[\chi]$ readily evaluate to (see Appendix~\ref{App:Qintegrals})
\bea
&&
B_i = 10 \frac{n_i}{T^3} \chi_i \ , \qquad
Q_{11}^{ij\to k\myell}
   = 30 (1+\delta_{k\myell}) 
       \frac{\sigma_{TOT}^{ij\to k\myell} n_i n_j}{T^4} \chi_i^2
 \ , \qquad 
Q_{21}^{ij\to k\myell}  = 0 \ \, \qquad
\nonumber \\ 
&&
Q_{31}^{ij\to k\myell}  
  = \frac{20}{3} (1+\delta_{k\myell}) 
    \frac{\sigma_{TOT}^{ij\to k\myell} n_i n_j}{T^4}\chi_i \chi_k \ , \qquad
Q_{41}^{ij\to k\myell}  
 = \frac{20}{3} (1+\delta_{k\myell}) 
    \frac{\sigma_{TOT}^{ij\to k\myell} n_i n_j}{T^4}\chi_i \chi_\ell \ ,
\eea
where we used (\ref{Q11_integration}), (\ref{Q21_integration}), 
and (\ref{Q31_integration}) with 
$E_a = p_a$ , $\gamma_3 = \beta_3 = 1/2$, 
and substituted
equilibrium densities
\be
n_i = \frac{g_i}{\pi^2} T^3  e^{\mu_i/T} \ .
\ee
For a one-component massless system,
\be
Q[\chi] = \frac{10 n}{T^3} \chi 
          \left (1 - \frac{10n \sigma_{TOT}} {3T} \chi\right) \ ,
\ee
which is maximal at $\chi^{Grad} = 3 T \lambda_{MFP}/20$, where
$\lambda_{MFP} = 1/n\sigma_{TOT}$ is the mean free path. So the viscous
correction is a dimensionless measure of the mean free path in this case.
The corresponding shear
viscosity from (\ref{eta_from_Q}) is the well-known Grad result
$\eta_s = 6T/5\sigma_{TOT}$.

\subsection{Two-component system in Grad approximation}
\label{Sc:Grad_2comp}

Extension to a minimalist multicomponent system with two massless species
and elastic two-body interactions involves three
interaction channels $A+A\to A+A$, $B+B\to B+B$, 
and $A+B \to A+B$. Crossing symmetry would also imply inelastic
$A+A\to B+B$ and $B+B \to A+A$ but these
are ignored here in order to isolate shear only
(if particle densities are allowed to change, there will also be dissipative 
effects due to particle diffusion). With
isotropic, energy-independent cross sections $\sigma_{AA}$, $\sigma_{BB}$, 
and $\sigma_{AB}$, for this system in Grad approximation,
\be
Q =\frac{10}{T}(n_A \chi_A  + n_B \chi_B)
    -\frac{100}{3T^4} (\sigma_{AA} n_A^2 \chi_A^2 +\sigma_{BB} n_B^2 \chi_B^2)
   + \frac{20\sigma_{AB}n_A n_B}{3T^4}(4\chi_A\chi_B - 7\chi_A^2 -7\chi_B^2) 
\ ,
\ee
which is maximized when
\bea
\chi_A^{Grad} &=& \frac{3LT}{20}
\frac{5K_{B(B)} + 7K_{B(A)} + 2K_{A(B)}}
              {K_{A(A)} [5 K_{B(B)} + 7 K_{B(A)}]
                    + K_{A(B)} [9K_{B(A)} + 7 K_{B(B)}]}
\nonumber\\
\chi_B^{Grad} &=&\frac{3LT}{20}
         \frac{5K_{A(A)} + 7K_{A(B)} + 2K_{B(A)}}
              {K_{B(B)} [5 K_{A(A)} + 7 K_{A(B)}]
               + K_{B(A)} [9 K_{A(B)} + 7 K_{A(A)}]} \ .
\label{cA_cB_massless}
\eea
Here $K_{i(j)} \equiv L / \lambda_{i(j)} = L n_j \sigma_{ij}$ 
denote partial inverse Knudsen numbers 
characterizing scattering of species $i$ off species $j$
and $L$ is the characteristic length scale for gradients in the system.
All four $K_{i(j)}$ play a role because
the solution to (\ref{chi_eq}) is influenced by any particle in the 
microscopic scattering process that is out of equilibrium 
(whether incoming, or outgoing).
The partial inverse Knudsen numbers also come with different weights, 
therefore, unlike for a single-component system, 
the result cannot in general 
be reproduced with just the mean free path as
$\chi_i \sim T\lambda_i \equiv LT/K_i = LT/\sum\limits_j K_{i(j)}$.
The Grad estimate of the shear viscosity
\be
\eta_s^{Grad} = \frac{6T}{5}
         \frac{\sigma_{AB} (7r + 7r^{-1} + 4) 
               + 5(\sigma_{AA}+\sigma_{BB})}
              {7 \sigma_{AB} (\sigma_{AA} r + \sigma_{BB} r^{-1})
              + 9\sigma_{AB}^2 + 5\sigma_{AA} \sigma_{BB}} 
\quad, \qquad r \equiv \frac{n_A}{n_B}
\ee
from (\ref{eta_from_Q})
is strictly speaking a variational lower bound on the exact $\eta_s$ value but 
usually reasonably accurate in practice (for the isotropic cross sections
used here).

\subsection{Comparison to nonlinear transport with 0+1D Bjorken expansion}
\label{Sc:BTE_2comp}

Linearized transport results correspond to the 
Navier-Stokes limit where the system
relaxed to a solution dictated by gradients of hydrodynamic variables. 
For expanding systems, such as those in heavy-ion collisions, 
relaxation to local equilibrium has to compete
with dilution and cooling, therefore it is important to check how well 
the limit applies when local equilibrium is no longer a static
fixed point in time. 

A convenient test scenario is a massless system 
undergoing boost-invariant 0+1D Bjorken expansion%
\footnote{
\label{Footnote:1}
By longitudinal boost invariance we mean that the state of the system
at each point in spacetime with $t>0$,  coordinate rapidity $\eta \ne 0$ 
can be obtained from the state on the $\eta = 0$ sheet via Lorentz
boost along the $z$ direction.
}
with homogeneous and isotropic transverse directions ($x,y$), 
just like in Ref.~\cite{isvstr0} 
but with a two-component $A+B$ mixture.
The system starts out at longitudinal proper time 
$\tau \equiv \sqrt{t^2 - z^2} = \tau_0$ in local thermal 
equilibrium but due to expansion
dissipative corrections quickly develop and can be easily quantified
using the partial shear stresses of the two species. 
Due to scaling of the transport 
solutions\cite{nonequil}
the evolution only depends on the dimensionless ratio 
$\ttau \equiv \tau / \tau_0$ and
partial inverse Knudsen numbers 
$K_{i(j)} \equiv \tau / \lambda_{i(j)} = \tau n_j \sigma_{ij}$,
where the characteristic scale for gradients is the proper time $\tau$.
The initial
temperature $T_0$ does not play any role beyond setting the momentum scale 
(all momenta are proportional to $T_0$).
As in Section~\ref{Sc:Grad_2comp}, we only include elastic two-body 
interactions $A+A\to A+A$, $B+B\to B+B$, and $A+B \to A+B$.
All three
cross sections are set to grow with time
as $\sigma_{ij} \propto \tau^{2/3}$, 
which ensures%
\footnote{
In a scale invariant system all cross sections are set by the temperature,
i.e., $\sigma \propto 1/T^2$.
However, as shown in Ref.~\cite{isvstr0}, 
$\sigma \propto 1/T^2$ 
is very well approximated by $\sigma \propto \tau^{2/3}$
because for 0+1D Bjorken expansion $T\propto \tau^{-1/3}$ as long as the
system is near local equilibrium.
}
approximately scale invariant dynamics
with $\eta_s/s \approx const$. In such a scenario,
longitudinal expansion first drives the system 
out of local equilibrium but at late times the system returns, 
asymptotically, to local equilibrium.

By symmetry, the phase space densities $f_i(\tau,p_T,\xi)$
only depend on proper time $\tau$,
transverse momentum magnitude $p_T$, and the difference $\xi \equiv \eta - y$ 
between coordinate rapidity $\eta$
and momentum rapidity $y$
(see Appendix~\ref{App:CooperFrye} for definitions).
The flow velocity is constrained
to $u^\mu = (\ch\, \eta, 0, 0, \sh\, \eta)$,
 and for both species shear stress is diagonal in the LR ($\eta = 0$) frame,
i.e.,
\be
\pi_{i,LR}^{\mu\nu} = diag(0, -\pi_{L,i}/2, -\pi_{L,i}/2, \pi_{L,i}) \ ,
\label{pimunu_LR_Bjorken}
\ee
where $\pi_{L,i}$ is the longitudinal shear stress for species $i$.
Assuming dissipative corrections are quadratic in momentum, we have
\be
\phi_i = c_i \frac{\pi^{\mu\nu}p_\mu p_\nu}{2(e+p)T^2}
         = c_i(\ttau) \frac{\pi_L(\ttau)}{8p(\ttau)} 
                       \frac{p_T^2}{T^2(\ttau)}\, 
                       \left(\sh^2\xi - \frac{1}{2}\right) 
 \quad \ \Rightarrow \qquad c_i(\ttau) 
  = \frac{\pi_{L,i}(\ttau)}{p_i(\ttau)} \frac{p(\ttau)}{\pi_L(\ttau)} \ ,
\label{df_Grad}
\ee
where $e_i = 3p_i$ was substituted for massless particles. 
Up to the factor $p/\pi_L$ that is common to all species,
$c_i$ describes how far species $i$ is from local equilibrium.
In the late-time Navier-Stokes regime, linearized kinetic theory 
predicts
\be
\frac{c_B}{c_A} = \frac{5 K_{A} + 2(K_{A(B)} + K_{B(A)})}
                       { 5K_B + 2(K_{A(B)}+K_{B(A)})} \qquad\qquad
(K_i \equiv \sum\limits_j K_{i(j)}) 
\label{linGrad_cA_cB}
\ee
(cf. (\ref{psi_with_c}) and (\ref{cA_cB_massless}), 
and note that the denominators in (\ref{cA_cB_massless}) cancel in the ratio).
The ``democratic Grad'' approach on the other hand postulates $c_i = 1$ for
all species, so $c_B / c_A = 1$.

\begin{figure}[h]
\includegraphics[width=0.49\linewidth]{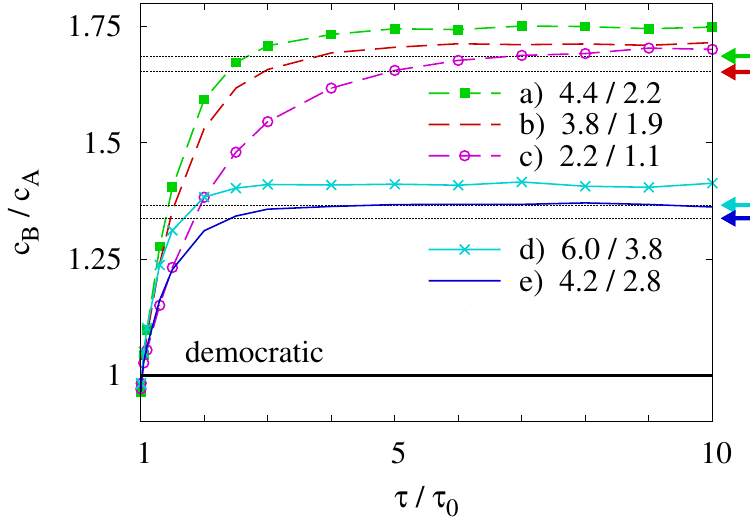}
\caption{Ratio of dissipative corrections as a function of normalized
proper time for a massless two-component
system in a 0+1D Bjorken scenario, calculated from
nonlinear $2\to 2$ covariant transport
using MPC\cite{MPC}. Five different scenarios a) - e) with various
cross sections and densities are shown,
labeled with the ratio of inverse Knudsen numbers $K_A / K_B$.
See Table~\ref{Table:cA_cB} for a detailed list of parameters.
Thin, horizontal dotted lines and arrows on the right side of the plot 
correspond to the expectation 
from a self-consistent calculation based on linearized
transport in the quadratic Grad approximation (``dynamical Grad'' approach).
Only four such lines and arrows are visible because
scenarios b) and c) are identical except for the timescale of relaxation to 
Navier-Stokes regime; scenario b) relaxes 5/3 times quicker than c).}
\label{Fig:cA_cB}
\end{figure}

Figure~\ref{Fig:cA_cB} compares these two extremes to fully nonlinear
transport solutions obtained using Molnar's Parton Cascade (MPC) 
\cite{MPC}. The simulations are initialized with 
uniform coordinate rapidity distributions $dN/d\eta$ 
in a wide window $|\eta| < 5$. To avoid 
the $|\eta| \gton 4$ edges of the system where boost invariance is strongly
violated,
shear stress evolution is extracted only using particles with
$|\eta|<2$ (all boosted to the $\eta = 0$ frame). 
A variety of relative cross sections and densities
between the two species are explored in five different scenarios shown 
in Table~\ref{Table:cA_cB}, 
which all keep species $A$ closer to equilibrium than $B$.
In all five cases, the ratio of viscous corrections $c_B/c_A$ starts from
unity but then relaxes to a 
constant value at late times that depends on the partial inverse 
Knudsen numbers
in the system. While the commonly used ``democratic Grad'' ansatz fails to
account for the species dependence of viscous corrections, 
linearized transport (Eq.~(\ref{linGrad_cA_cB})) captures the corrections 
with better than 10\% accuracy in all five scenarios despite rapid
longitudinal expansion.

\begin{table}
\caption{Inverse Knudsen numbers for the two species, 
and ratios of densities and cross sections,
for the two-component massless covariant transport calculation
in Fig.~\ref{Fig:cA_cB}.}
\label{Table:cA_cB}
\begin{center}
\begin{tabular}{c c c c c}
\hline \hline
Scenario & $K_A$ & $K_B$ & 
$n_A : n_B$ & $\sigma_{AA} : \sigma_{AB} : \sigma_{BB}$ \cr
\hline
a) & 4.4 & 2.2 & 3 : 1 & 20 : 10   :  5 \cr
b) & 3.8 & 1.9 & 2 : 2 & 20 : 10   :  5 \cr
c) & 2.2 & 1.1 & 2 : 2 & 12 :  6   :  3 \cr
d) & 6   & 3.8 & 1 : 3 & 24 : 24   : 12 \cr
e) & 4.2 & 2.8 & 2 : 2 & 20 : 13.3 : 8.89 \cr
\hline \hline
\end{tabular}
\end{center}
\end{table}

Shear stress evolution in a particle mixture has also been studied in \cite{El:2012ka}, albeit
using a different approach based on imposing 
the second law of thermodynamics (entropy production).
In that work, an approximate relation for partial shear stress ratios
has also been obtained (cf. (12) therein).
While that result qualitatively captures both the rise and saturation of the curves 
in Fig.~\ref{Fig:cA_cB}, quantitatively, the predicted asymptotic values 
are not identical to (\ref{linGrad_cA_cB}) here.
However, those results are for an assumed 
uniform flow velocity across the entire system, which is
inconsistent with $\pi^{\mu\nu} \sim \nabla^\mu u^\nu$ in 
the Navier-Stokes regime considered here.
It would be interesting to compare these two approaches in more detail in the future.

\section{Massive two-component system}
\label{Sc:massive_2comp}

For nonrelativistic particles, in the Grad approximation 
(see Appendix~\ref{App:Grad_NR}),
\bea
B_i &=& \frac{5z_i}{2}\frac{n_i}{T^3} \chi_i
\nonumber \\
Q_{11}^{ij\to k\myell}
   &=& \frac{1}{3\sqrt{2\pi}} \frac{z_i^{3/2}}{z_j^{1/2}} 
       \frac{15z_i^2 + 40 z_i z_j + 24 z_j^2}{(z_i + z_j)^{3/2}}
       (1+\delta_{k\myell}) 
       \frac{\sigma_{TOT}^{ij\to k\myell} n_i n_j}{T^4} \chi_i^2
\nonumber \\
Q_{21}^{ij\to k\myell} 
   &=& -\frac{1}{3\sqrt{2\pi}}
         \left(\frac{z_i z_j}{z_i + z_j}\right)^{3/2} (1+\delta_{k\myell}) 
         \frac{\sigma_{TOT}^{ij\to k\myell} n_i n_j}{T^4} \chi_i \chi_j
\nonumber \\ 
Q_{31}^{ij\to k\myell}  
  &=& \frac{5}{\sqrt{2\pi}}\frac{z_i^{3/2} z_k^2}{z_j^{1/2} (z_i + z_j)^{3/2}}
    (1+\delta_{k\myell}) 
    \frac{\sigma_{TOT}^{ij\to k\myell} n_i n_j}{T^4}\chi_i \chi_k
\nonumber\\
Q_{41}^{ij\to k\myell}  
 &=& \frac{5}{\sqrt{2\pi}}
     \frac{z_i^{3/2} z_\myell^2}{z_j^{1/2} (z_i + z_j)^{3/2}}
     (1+\delta_{k\myell}) 
    \frac{\sigma_{TOT}^{ij\to k\myell} n_i n_j}{T^4}\chi_i \chi_\myell
\label{Qvalues_NR}
\eea
where $z\equiv m/T$ and 
equilibrium densities
\be
n_i^{NR} = \frac{g_i}{(2\pi)^{3/2}} (m_i T)^{3/2}  e^{(\mu_i - m_i)/T}
\ee
were substituted.
For a one-component nonrelativistic system, the above imply
\be
\chi^{Grad} = \frac{5\sqrt{\pi}}{32}\sqrt{\frac{T}{m}} \frac{T}{n\sigma_{TOT}} 
\quad
\Rightarrow \quad
\eta_s^{Grad} = \frac{5\sqrt{\pi}}{16} \frac{\sqrt{m T}}{\sigma_{TOT}} \ ,
\label{etas_NR_Grad}
\ee
reproducing the familiar nonrelativistic viscosity expression.
Notice that for fixed density and cross section the relative
viscous correction $\delta f/\feq$ 
{\em decreases} when mass increases, even though shear 
viscosity increases with mass.

For a one-component system the shear 
viscosity is known analytically, in Grad approximation,
for arbitrary $m/T$ with fully relativistic kinematics 
(see Chapter XI of Ref.~\cite{deGroot}):
\be
\eta_s^{Grad} = \frac{15 z^2 K_2^2(z) h^2(z)}
               {16[(15 z^2 + 2) K_2(2z) + (3z^3+49z) K_3(2z)]} 
\, \frac{T}{\sigma_{TOT}} \qquad \ , \qquad h(z) \equiv z K_3(z) / K_2(z) \ ,
\label{etas_Grad}
\ee
where $K_n$ is a modified Bessel function of the second kind.
The numerical integration method in Appendix~\ref{App:Qintegrals}
reproduces this result,
and we also rechecked the complete derivation of the formula
in Ref.~\cite{deGroot}
(note the typographic error in the book;
 the correct coefficient in the denominator is 15, not 5).

\subsection{Two-component nonrelativistic system in Grad approximation}

For a two-component nonrelativistic 
$A+B$ system with isotropic, energy-independent, elastic
scattering, in Grad approximation
\be
Q[\chi_A,\chi_B]
      = \left[   \frac{5 z_A  n_A \chi_A}{2T^3} 
               - \frac{8\sigma_{AA} n_A^2 z_A^{3/2} \chi_A^2}
                              {\sqrt{\pi}\,T^4}
               +\frac{8\sqrt{2}\,\sigma_{AB} n_A n_B z_A^{3/2} z_B^{1/2} 
                       [(5z_A + 3z_B) \chi_A - 2z_B \chi_B] \chi_A}
                     {3\sqrt{\pi}\,T^4(z_A+z_B)^{3/2}}
        \right]
        + A\leftrightarrow B \ .
\ee
The general structure of the solution is very similar to the massless case,
namely, all partial inverse Knudsen numbers contribute with different
weights that now also
depend on the masses. In the limit when species $B$ is much more dilute
than species $A$ (for example, because it is very heavy), 
we can approximate $n_B \to 0$ to obtain
\bea
\left.\chi_A^{Grad}\right|_{n_B\to 0} 
      &=& \frac{5\sqrt{\pi}}{32}\sqrt{\frac{T}{m_A}}\frac{T}{\sigma_{AA} n_A} 
   \quad , \qquad
\nonumber\\
\left.\chi_B^{Grad}\right|_{n_B\to 0} 
      &=& \chi_A^{Grad} 
          \frac{3(\mu+1)^2 \sigma_{AA} + 2\sqrt{2\mu(1+\mu)} \sigma_{AB}}
               {\sqrt{2\mu(1+\mu)} (3+5\mu) \sigma_{AB}} \qquad\quad
               ( \mu = \frac{m_B}{m_A}) \ .
\label{chiRatio_NR_nB0}
\eea
In this special case species $A$ is unaffected by species $B$,
and also $\sigma_{BB}$ is irrelevant.
On the other hand, for species $B$ we have
\bea
\frac{\chi_B}{\chi_A} &=& \frac{3\sigma_{AA}}{4\sigma_{AB}} + \frac{1}{4}
\qquad \qquad {\rm if}\quad m_A = m_B,
\nonumber\\
\frac{\chi_B}{\chi_A} &\approx& \frac{3\sigma_{AA}}{5\sqrt{2}\, \sigma_{AB}}
\qquad\qquad {\rm if} \quad m_B \gg m_A \ ,
\eea
which tells that the heavier species tends to have {\em smaller} 
viscous correction
even when its interaction cross section is the same as that of the light species.

\subsection{Pion-nucleon gas and elliptic flow}
\label{Sc:piN}

Next consider a more realistic pion-nucleon system, with relativistic 
kinematics.
Lumping isospin states and 
antiparticles into a single species, this is a two-component system
with $m_\pi = 0.14$ GeV, $g_\pi = 3$, $m_N = 0.94$ GeV, $g_N = 4$. For
temperatures $120$~MeV $\lton T \lton 165$~MeV of interest 
we approximate the two-body
cross sections with constant, energy-independent, effective values 
$\sigma^{eff}_{\pi\pi} = 30$~mb, $\sigma^{eff}_{\pi N} = 50$~mb, 
and $\sigma^{eff}_{NN} = 20$~mb.
These values are set so that for a static system ($u^\mu = (1,\vzero)$) 
in thermal and
chemical equilibrium the mean times $\bar\tau_{i(j)}$ between scatterings
for particles of species $i$ with particles of species $j$, defined through
\be
\frac{1}{\bar\tau_{i(j)}} = \langle n_j\sigma_{ij} v_{rel}\rangle = 
\frac{1}{n_i} 
  \int \frac{d^3 p_1}{E_1} \frac{d^3 p_2}{E_2} \feq_i(\vp_1) \feq_j(\vp_2)
           \sigma_{ij} F(s) \ ,
\label{tauij_inv}
\ee
are comparable to the values shown in Figs.~2b and 5a of 
Ref.~\cite{Prakash} (Table~\ref{Table:gamma_piN} lists the mean scattering
times with these effective cross sections as a function of temperature,
including $T=100$ and 200~MeV outside the matching range). Here
\be
F(s) \equiv p_{cm} \sqrt{s} \equiv E_1 E_2 v_{rel} 
= \frac{1}{2}\sqrt{(s - m_i^2 - m_j^2)^2 - 4 m_i^2 m_j^2}
\label{Fs}
\ee
is the flux factor. 
Note that at 
these temperatures pions are much more abundant than
nucleons, and therefore nucleon-nucleon scattering affects 
viscous corrections negligibly 
(one could put $\sigma_{NN} = 0$ to good approximation).

\begin{table}
\caption{Mean scattering times in a pion-nucleon gas with
effective cross sections $\sigma_{\pi\pi}^{eff} = 30$~mb, $\sigma_{\pi N}^{eff} = 50$~mb,
and $\sigma_{N N}^{eff} = 20$~mb. 
Values are rounded to the two most significant 
digits.}
\label{Table:gamma_piN}
\begin{center}
\begin{tabular}{c c c c}
\hline \hline
$T$~[MeV] & $\bar\tau_{\pi\pi}$ [fm] &
$\bar\tau_{N(\pi)}$ [fm] & $\bar\tau_{NN}$ [fm]  \cr
\hline
100 & 12.7 & 8.2 & 8300 \cr
120 &  6.6 & 4.2 & 1200 \cr
140 &  3.9 & 2.4 & 280 \cr
165 &  2.2 & 1.4 & 73 \cr
200 &  1.2 & 0.73 & 18 \cr
\hline \hline
\end{tabular}
\end{center}
\end{table}

\begin{figure}[h]
\includegraphics[width=0.49\linewidth]{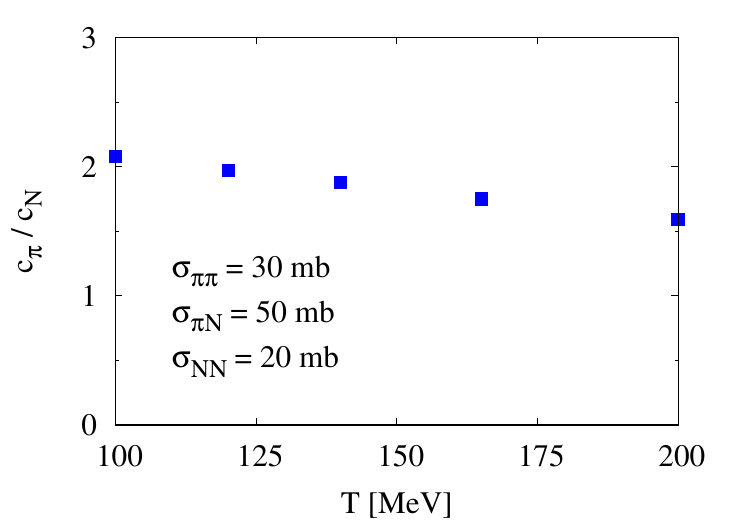}
\caption{Self-consistent dissipative corrections for shear stress
as a function of temperature for a chemically equilibrated pion-nucleon gas,
in the Grad approximation,
with effective cross sections $\sigma_{\pi\pi}^{eff} = 30$~mb, 
$\sigma_{\pi N}^{eff} = 50$~mb, and
$\sigma_{NN}^{eff} = 20$~mb. The ratio of coefficients $c_\pi / c_N$ is shown,
where $c_i$ is the dissipative correction for species $i$
{\em relative} to the commonly used ``democratic'' ansatz (see text).}
\label{Fig:pi_N_coeffs}
\end{figure}

For the $\pi-N$ system, 
the ratio of viscous coefficients is $c_\pi / c_N \sim 2$
in the temperature window $100 < T < 200$~MeV, 
as shown in Figure~\ref{Fig:pi_N_coeffs}. This means that nucleons are 
about twice as close as pions to equilibrium
(at the same momentum), in qualitative agreement
with the analytic results in Section~\ref{Sc:massive_2comp}.
For example, the nonrelativistic formula (\ref{chiRatio_NR_nB0}) would predict 
$c_\pi / c_N \approx 2.9$, which is not bad considering that pions
are relativistic at these temperatures.
The primary origin of the pion-nucleon difference is the
larger $\pi N$ cross section -- a nucleon scatters more frequently off pions
than a pion scatters off another pion. But based on the earlier discussion 
one would expect $c_\pi > c_N$ even for $\sigma_{\pi\pi} = \sigma_{\pi N}$.

The above pion-nucleon difference is reflected in pion vs proton 
observables if the self-consistent, species-dependent viscous corrections 
are included in Cooper-Frye freezeout. 
To estimate the effect, we perform a hydrodynamic simulation of $Au+Au$ at 
top RHIC energy $\sqrt{s_{NN}}=200$~GeV with impact parameter $b=7$~fm,
and look at the difference between pion and proton elliptic flow. 
The calculations
are done with {\sf AZHYDRO}~\cite{AZHYDROrefs,AZHYDRO} version 0.2p2,
which is a 2+1D code with longitudinal boost invariance. 
This version includes the fairly recent {\sf s95-p1} equation of state
parameterization\cite{EOSs95p1} 
by Huovinen and Petreczky that matches lattice QCD results to a hadron
resonance gas.
Because there is no
dissipation in {\sf AZHYDRO}, we estimate shear stress
on the conversion hypersurface from gradients of the ideal flow 
fields using the Navier-Stokes formula (\ref{trcoeff_def}), 
i.e., $\pi^{\mu\nu} = \eta_s \sigma^{\mu\nu}$. 
This is in the same spirit as an early exploration of shear stress
corrections by Teaney~\cite{Derek_shear}, 
except we use real hydrodynamic solutions instead of a parameterization.
We set $\eta_s/s = 0.1$, and determine the shear viscosity from
the hydrodynamic solutions using 
\be
  \eta_s = \frac{\eta_s}{s} \frac{e+p}{T} \qquad\qquad (\mu_B = 0) \ .
\ee
For initial conditions at Bjorken proper time $\tau_0 = 0.5$~fm we set
the transverse entropy density distribution $ds/d^2x_T d\eta$ to
a 25\%+75\% weighted sum of binary collision and wounded nucleon profiles
($\sigma_{NN}^{inel} = 40$~mb),
with diffuse Woods-Saxon nuclear densities for gold nuclei 
(Woods-Saxon parameters
$R=6.37$~fm, $\delta=0.54$~fm), a peak entropy density value 
$s_0 = \frac{1}{\tau_0} \frac{ds(\vx_T = 0)}{d^2 x_Td\eta} = 110$/fm$^3$, 
and vanishing baryon density $n_B = 0$ everywhere. 
With ordinary, ideal ($\delta f =0$) 
Cooper-Frye freezeout at temperature $T_{conv} = 140$~MeV, 
this roughly reproduces
the measured pion spectrum.
In the following we keep the initial conditions fixed but vary
$T_{conv}$, and study pion and proton elliptic flow 
from fluid-to-particle conversion 
with self-consistent viscous $\delta f$ 
corrections. The viscous Cooper-Frye procedure is discussed in
Appendix~\ref{App:CooperFrye} 
(the {\sf AZHYDRO} code only handles ideal freezeout, i.e., $\delta f = 0$).

\begin{figure}[h]
\includegraphics[width=0.49\linewidth]{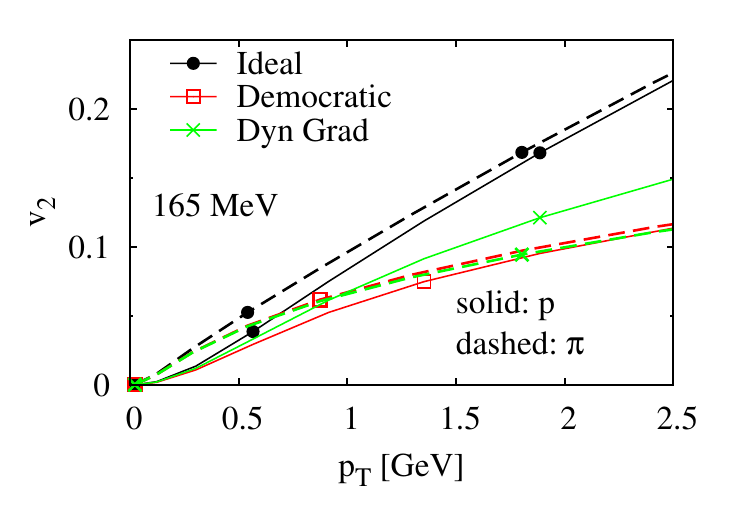}
\includegraphics[width=0.49\linewidth]{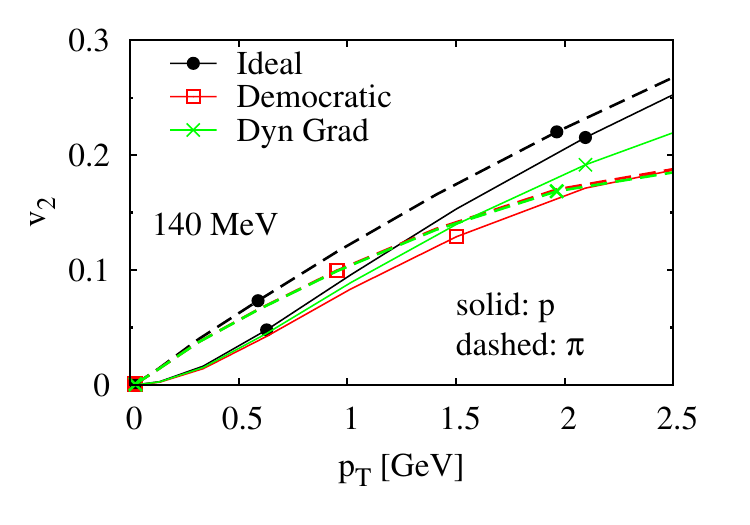}
\caption{Differential elliptic flow $v_2(p_T)$ 
of pions and protons in $Au+Au$ at 
$\sqrt{s_{NN}}=200$~GeV at RHIC with impact parameter $b=7$~fm, using
2+1D boost invariant hydrodynamic solutions from 
AZHYDRO\cite{AZHYDROrefs,AZHYDRO}, and Cooper-Frye fluid-to-particle
conversion at $T_{conv} = 165$~MeV (left plot) or 140~MeV (right plot).
Dashed lines are for pions, while solid curves are for protons.
The standard ``democratic Grad'' approach (open boxes) is compared
to self-consistent shear corrections (crosses) 
computed for a pion-nucleon gas from
linearized kinetic theory (see text). In both cases,
$\eta_s/s = 0.1$ at conversion. Results with uncorrected, local equilibrium
phase space distributions ($\delta f = 0$) are also shown (filled
circles). 
}
\label{Fig:pi_p_v2pt_165-140}
\end{figure}

The left plot in Figure~\ref{Fig:pi_p_v2pt_165-140} shows 
differential elliptic flow results for pions and protons 
for freezeout at $T_{conv} = 165$~MeV.
Pion and proton $v_2$ separate already in the ideal case (filled circles),
following the characteristic mass ordering of $v_2$ in hydro.
At high $p_T$ this
effect diminishes, however. 
Viscous freezeout with the commonly used democratic ansatz (open boxes)
preserves the
mass ordering but with $v_2$ strongly suppressed by dissipation, even
for the modest $\eta_s/s = 0.1$ used here.
In this calculation dissipative effects are only present 
in the viscous phase space
corrections $\delta f_i$ at fluid-to-particle conversion but viscous
corrections to the evolution of hydrodynamic flow and temperature fields
are known\cite{teaneyv2,IStrv2,Monnai:2009ad} 
to have smaller influence on $v_2$ than $\delta f$ itself.
In contrast, self-consistent species-dependent freezeout (crosses) 
leads to a clear pion-proton elliptic flow splitting at moderately high
transverse momenta, with the proton $v_2$ exceeding the pion $v_2$ by
$30$\%.
Both species exhibit a strong viscous suppression in $v_2$. 
However, the suppression is smaller for protons because they are more
equilibrated than pions. At low $p_T$ the mass effect is still present,
which means that the pion and proton elliptic flow curves necessarily cross
each other (at around $p_T \sim 1$ GeV in this calculation). 
The reason why the pion results are almost identical to ``democratic'' 
freezeout is that at $T=165$~MeV the pion density is much higher than the 
proton density, i.e., the dynamics of pions is largely unaffected by
the protons, and both the shear viscosity and the entropy density are 
then dominated by pions. The temperature $T=165$~MeV used here is 
the same as the typical switching temperature in hybrid hydro+transport 
models\cite{VISHNU}. 
It would be very interesting to initialize the transport
stage of hybrid calculations 
with self-consistent viscous distributions for each species,
and check the effect on identified particle 
elliptic flow at the end of the hadron transport evolution.

The right plot of Fig.~\ref{Fig:pi_p_v2pt_165-140} 
shows the same $v_2(p_T)$ calculation but with a lower $T_{conv} = 140$~MeV.
The qualitative picture is the same, but in
this case the viscous suppression of $v_2$ is smaller in magnitude
because, for the Navier-Stokes stresses (\ref{trcoeff_def}) 
used here, flow gradients 
$\partial^{\mu} u^\nu \sim 1/\tau$ are smaller. The mass ordering is also
stronger, which is expected because it is driven by $m/T$. At the same
$p_T \sim 2-2.5$~GeV, the relative difference between proton $v_2$ curves
from the ``democratic'' and the self-consistent approaches is smaller
than for $T_{conv}=165$~MeV.
However,
the relative {\em change} in viscous suppression of $v_2$ 
is actually larger; the difference for protons between ideal hydrodynamic
freezeout and the viscous result shrinks by a factor of two at 
$T_{conv}=140$~MeV 
when the fluid is converted to particles with the self-consistent 
(species-dependent) scheme.

At even lower temperature $T_{conv} = 120$~MeV, dissipative corrections for 
$\eta_s/s = 0.1$ are basically negligible for
protons for $p_T < 2.5$~GeV, at least with the Navier-Stokes shear stress used here.  For pions there is a less then 10\% suppression
in $v_2$ at high $p_T$.

\subsection{Simple four-source model of viscous elliptic flow}
\label{Sc:foursource}

The elliptic flow results presented in Sec.~\ref{Sc:piN} come from
numerical hydrodynamic solutions, where both inhomogeneities over the Cooper-Frye hypersurface
and also the shape of the hypersurface matter. It is desirable to gain
at least some qualitative analytic 
insight into how viscous corrections affect differential
$v_2(p_T)$ for particles of different masses. To this end we
generalize the simple model in \cite{Huovinen:2001cy} (cf. Fig.~6 therein),
which considered four uniform, non-expanding fireballs 
boosted symmetrically in back-to-back pairs along the $x$
and $y$ directions in the transverse plane, respectively, 
with velocities $\pm v_x$ and $\pm v_y$ ($v_x > v_y \ge 0$). 
All four sources have the same temperature, chemical potential,
and volume in the laboratory frame. Isochronous $t = const$ emission 
is considered at zero momentum rapidity, in which case flow coefficients
are given by
\be
v_n(p_T) 
= \frac{ \int\limits_0^{2\pi} d\phi\, f(p_T, \phi) \, \cos (n\phi) }
       { \int\limits_0^{2\pi} d\phi\, f(p_T, \phi) }
\label{def_vnpt}
\ee
with%
\footnote{
We trust that the reader will not confuse the momentum rapidity 
variable $y$ and the $y$ axis in the transverse plane.
}
\be
 f(p_T,\phi) \equiv f_{(+x)}(p_T, \phi, y=0) + f_{(-x)}(p_T, \phi, y=0)
                    + f_{(+y)}(p_T, \phi, y=0) + f_{(-y)}(p_T, \phi, y=0) \ .
\ee
For each source we take viscous corrections of the Grad form (\ref{df_Grad})
with uniform shear stress across the fireball, where $\pi^{\mu\nu}$ is a 
boosted copy of the 0+1D Bjorken shear stress solution 
(\ref{pimunu_LR_Bjorken}).
This way, the
dimensionless $\kappa \equiv \pi_L / (e+p)$ is our only extra parameter,
and it is the same for all four fireballs. Note that 
for typical viscous 0+1D Bjorken evolution $\kappa < 0$ because the 
longitudinal shear stress is negative 
(see \cite{isvstr0} for an extensive analysis).
Setting $\kappa = 0$ 
reproduces the ideal fluid results in~\cite{Huovinen:2001cy}. 

\begin{figure}[h]
\includegraphics[width=0.49\linewidth]{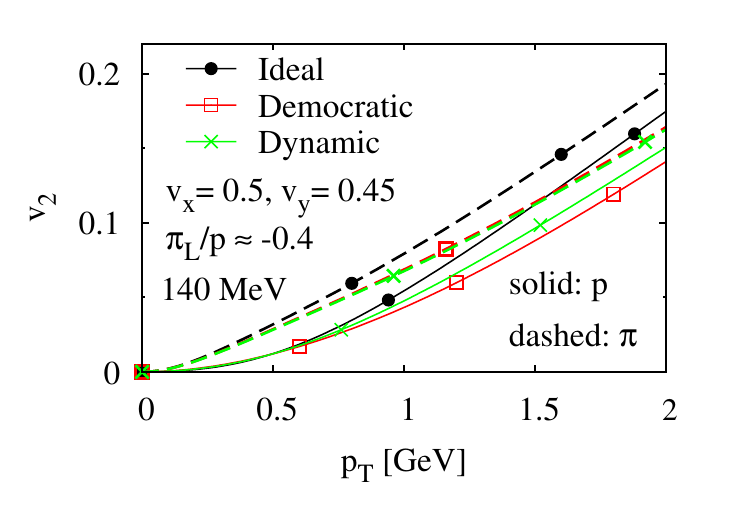}
\caption{Differential elliptic flow $v_2(p_T)$ from a viscous
generalization of the simple four-source model of Ref.~\cite{Huovinen:2001cy}
with parameters $v_x = 0.5$, $v_y = 0.45$, $T = 140$~MeV, 
and $\kappa \equiv \pi_L/(e+p) = -0.06$.
Dashed lines are for pions, while solid curves are for protons.
The standard ``democratic Grad'' approach
(open boxes, $c_\pi = c_p = 1$) is compared
to self-consistent shear corrections (crosses, $c_\pi = 1.03$ and $c_p = 0.55$) computed for a pion-nucleon gas from linearized kinetic theory (see Sec.~\ref{Sc:piN}). 
Results with uncorrected, local equilibrium
phase space distributions ($c_\pi = c_p = 0$) are also shown (filled circles). 
}
\label{Fig:v2_4source}
\end{figure}

Straightforward calculation yields anisotropic flow coefficients 
in terms of modified Bessel functions of the first kind (see App.~\ref{App:foursource} for details). Here we only discuss $v_2$ but in general
all even $v_n$ are nonzero. From (\ref{res_vnpt}),
\be
v_2(p_T) = \frac{G_2(a_x,b_x,z,c\kappa) - G_2(a_y,b_y,z,c\kappa)}
                {G_0(a_x,b_x,z,c\kappa) + G_0(a_y,b_y,z,c\kappa)} \ ,
\ee
where
\be
z \equiv \frac{m}{T} \ ,
\quad
a_i \equiv \frac{\sqrt{m^2+p_T^2}}{T \sqrt{1-v_i^2}} \ ,
\quad
b_i \equiv \frac{v_i p_T}{T \sqrt{1-v_i^2}}
\qquad (i = x, y) \ ,
\ee
$c$ is the magnitude of the viscous correction 
{\em relative} to the ``democratic'' Grad case, 
and $G_n$ is given by (\ref{def_G}).
Figure~\ref{Fig:v2_4source} 
shows the result for $T=140$~MeV, 
$v_x = 0.5$, $v_y = 0.45$, with $\kappa = -0.06$ which corresponds
to $\pi_L/p \approx -0.4$ (at this temperature $e/p \approx 5.5$).
The local thermal equilibrium curves (filled circles) exhibit the 
well-known mass ordering of ellipit flow  
(see \cite{Huovinen:2001cy} for more discussion).
Relative to this baseline, viscous corrections reduce elliptic
flow for both protons and pions. The reduction for protons, however,
is only half as large from the self-consistent approach (crosses) 
compared to the ``democratic'' Grad ansatz (open boxes).
Though this simple model does not capture 
the flattening of $v_2$ at  $p_T \gton 1.5$~GeV in Fig.~\ref{Fig:pi_p_v2pt_165-140}, it does illustrate that viscous corrections generally make 
elliptic flow smaller.

\section{Multicomponent hadron gas}
\label{Sc:multi}

In Section~\ref{Sc:piN} self-consistent corrections were calculated 
for a pion-nucleon gas. This is clearly only an estimate because it ignores
interactions of pions and nucleons with other species in the system.
It is natural to extend the investigation to  
mixtures with many hadronic species, in which case each species will have
its own dissipative corrections based on the microscopic dynamics.
The problem is complicated, however,
because it requires knowledge of hadronic scattering rates
between all species. In principle these are encoded in hadron transport
codes, such as UrQMD\cite{UrQMD}, AMPT\cite{AMPT}, or JAM\cite{JAM},
and we plan to apply these in a future study. 
Here we only pursue two simple models: 
i)
a hadron gas with the same, fixed scattering cross section for all 
species, which is the model in Ref.~\cite{DenicolQM2012};
and ii) a gas with more realistic
cross sections that follow additive quark
model\cite{AQM,UrQMD} (AQM) scaling, 
i.e., constant meson-meson, meson-baryon, and 
baryon-baryon cross sections with ratios 
$\sigma_{MM} : \sigma_{MB} : \sigma_{BB} = 4:6:9$. In both cases
we only consider elastic $ij\to ij$ channels (allowing for $i=j$), 
with energy-independent, isotropic
cross sections.

For the fixed cross section scenario we use $\sigma_{ij} = 30$~mb, the same
value as the effective $\sigma_{\pi\pi}$ for the pion-nucleon gas earlier
(cf. Fig.~\ref{Fig:pi_p_v2pt_165-140}).
For the AQM model, we take $\sigma_{MM} = 30$~mb, which implies $\sigma_{MB}=45$~mb, and $\sigma_{BB} = 67.5$~mb. 
To simplify the computation, 
we combine, as in Section~\ref{Sc:piN}, members of the same isospin multiplet,
and their antiparticle partners as well, into a single 
species with appropriately scaled degeneracy so that the number of degrees of
freedom and the particle densities stay the same.
The following calculation includes hadrons up to $m=1.672$ GeV, i.e., 
the $\Omega(1672)$, which 
translates into 49 effective species (the $c_i$ coefficients for the
49 species in the various scenarios are listed in Appendix~\ref{App:tables}).

\subsection{Elliptic flow for mixture in Grad approximation}
\label{Sc:species49_Grad}

\begin{figure}[h]
\includegraphics[width=0.49\linewidth]{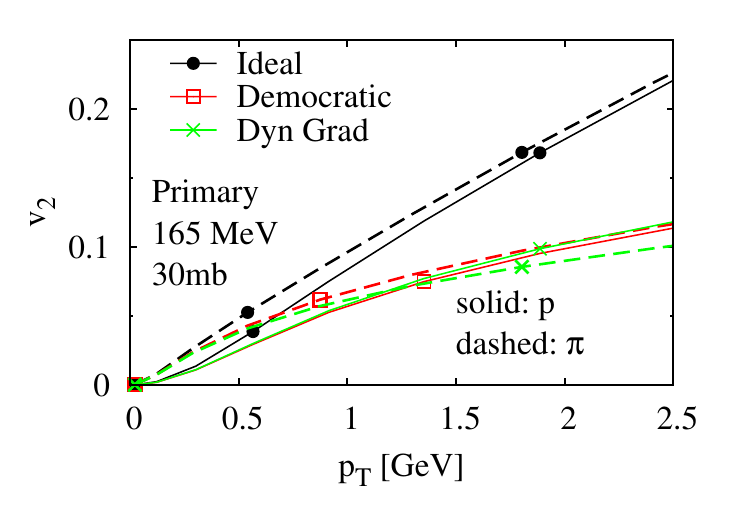}
\includegraphics[width=0.49\linewidth]{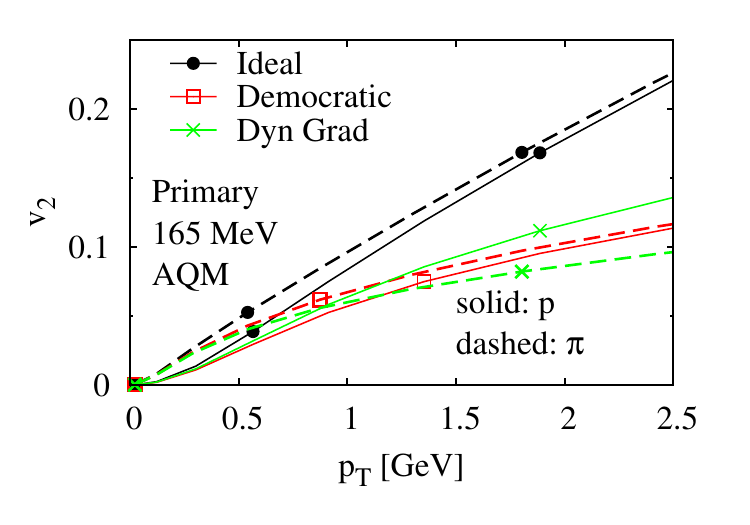}
\caption{Same as Fig.~\ref{Fig:pi_p_v2pt_165-140}, except the self-consistent
viscous corrections are computed for a gas of all hadron species
up to m=1.672~GeV ($\Omega(1672)$), 
with members of each isospin multiplet (and antiparticles) combined together 
into a single effective species. There are 49 effective species this way. 
{\em Left plot:} all hadron species interacting with the 
same constant isotropic cross section $\sigma_{ij}=30$~mb. 
{\em Right plot:} constant isotropic
cross sections with additive quark model scaling 
$\sigma_{MM} : \sigma_{MB} : \sigma_{BB} = 4:6:9$ and $\sigma_{MM} = 30$~mb.
Calculations with the ``democratic Grad'' ansatz for $\eta_s/s = 0.1$ 
(open boxes) and with local
equilibrium distribution (filled circles) are also shown.
In all cases, and for both plots,
the Cooper-Frye prescription is applied at $T_{conv}=165$~MeV.
}
\label{Fig:HG165_v2pt_primary}
\end{figure}

Figure~\ref{Fig:HG165_v2pt_primary} shows pion and proton elliptic flow 
$v_2(p_T)$
in $Au+Au$ at RHIC at $b=7$~fm from a calculation analogous to the $\pi-N$ 
system in 
Section~\ref{Sc:piN} with Cooper-Frye particle conversion
applied at $T_{conv}=165$~MeV, 
except now with self-consistent phase space corrections
$\delta f_i$ calculated for the multicomponent hadron gas.
The left plot is for $\sigma_{ij} = const$, in which
case pion and proton elliptic flow are very close to results from 
the ``democratic''
approach. The lack of species dependence is very similar to the 
findings of Ref.~\cite{DenicolQM2012}. 
If one looks closely, however, at high $p_T$, 
proton flow is actually slightly higher than pion 
flow, reflecting the decrease in 
shear stress corrections with mass 
 at fixed cross section (cf. Section~\ref{Sc:massive_2comp}).

The right plot of Fig.~\ref{Fig:HG165_v2pt_primary} shows, on the other hand, 
that more realistic additive quark model cross sections do generate
a pion-proton difference in elliptic flow, of similar magnitude to 
the difference seen for a pion-nucleon gas earlier. Crossing between pion
and proton $v_2$ also happens at about the same $p_T \sim 1$~GeV.
The likely explanation for this is that even though interactions with all
species are now considered, interactions with pions dominate because at 
$T_{conv}=165$~MeV pions have a 
much higher density compared to all other species, including kaons,
the second lightest species.
Though not shown here, we note that for $T_{conv}=140$~MeV one finds the same:
the fixed cross section scenario closely matches
the ``democratic'' Grad results, whereas pion-proton splitting in the AQM
scenario is very similar in magnitude to the $T_{conv}=140$~MeV 
result of Fig.~\ref{Fig:pi_p_v2pt_165-140} (right plot).

\begin{figure}[h]
\includegraphics[width=0.49\linewidth]{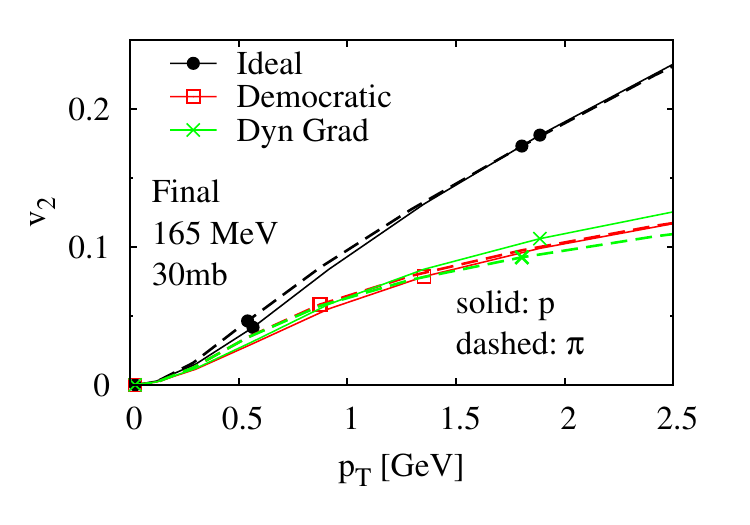}
\includegraphics[width=0.49\linewidth]{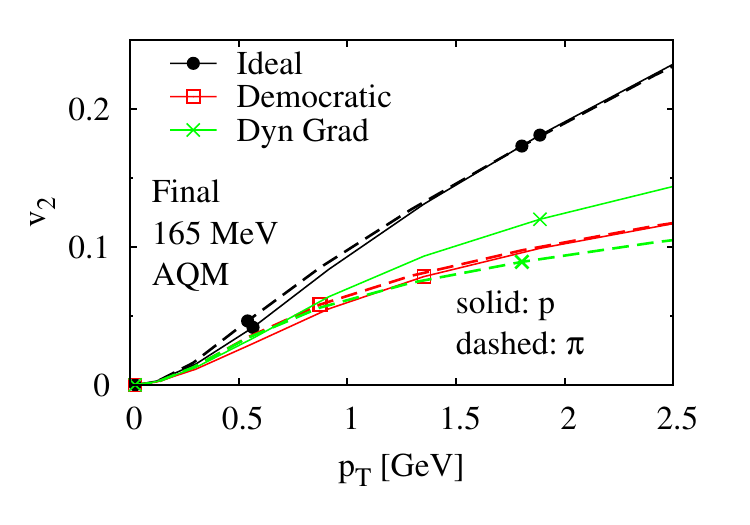}
\caption{Same as Fig.~\ref{Fig:HG165_v2pt_primary}, 
except after feeddown from resonance decays using the {\sf RESO} code in the 
{\sf AZHYDRO} package\cite{AZHYDROrefs}.}
\label{Fig:HG165_v2pt_decays}
\end{figure}

The Cooper-Frye prescription gives the momentum distribution of particles
emitted directly from the fluid (``primary'' particles). In a pure
hydrodynamic approach, i.e., without a hadronic afterburner, 
many of these
particles later decay en route to the detectors. 
Figure~\ref{Fig:HG165_v2pt_decays} shows the $p_T$ dependence of
pion and proton elliptic flow from the same calculation shown in 
Fig.~\ref{Fig:HG165_v2pt_primary},
except unstable resonances are decayed using the {\sf RESO} code in the 
{\sf AZHYDRO} package\cite{AZHYDROrefs} 
(stable hadrons in {\sf RESO} are the pions, kaons, and nucleons).
For ideal freezeout ($\delta f = 0$), the ``democratic Grad'' ansatz,
and also the constant cross section scenario,
the main effect of resonance decays on elliptic flow is 
a reduction of the pion-proton
splitting at low $p_T$, while at high $p_T$ there is barely any effect.
At $T_{conv} = 165$~MeV the difference between
pions and protons for all three scenarios gets washed out almost completely
(this is not universal at all temperatures,
for lower $T_{conv} = 140$ or 120~MeV, 
 a portion of the difference survives).
In contrast, in the more realistic AQM scenario, with 
self-consistent viscous fluid-to-particle conversion, 
proton
elliptic flow stays 30\% higher at $p_T \sim 2$~GeV 
than pion elliptic flow even after resonance decays are taken into account.
The same insensitivity to resonance decays is present 
at $T_{conv} = 140$~MeV and 120~MeV as well (not shown).

\subsection{Elliptic flow for mixture with {\boldmath $\delta f\propto p$ 
 or $p^{3/2}$}}
\label{Sc:species49_pto32}

Finally to investigate systematic errors due to the assumed quadratic momentum
dependence of dissipative corrections (Grad ansatz),
we explore instead power law momentum dependence with
$\delta f_i/\feq_i \propto p$ and $p^{3/2}$. These
correspond to (\ref{chi_def}) with
\bea
\chi_i^{(1)}(|\tilde\vp|) &=& c_i |\tilde\vp|^{-1} \ \chi_i^{dem} \nonumber \\
\chi_i^{(3/2)}(|\tilde\vp|) &=& c_i |\tilde\vp|^{-1/2} \ \chi_i^{dem} \nonumber \\
\qquad\qquad (\chi_i^{Grad} &=& c_i |\tilde\vp|^0) \ ,
\label{df_corrections}
\eea
where coefficients are determined variationally via maximizing $Q[\chi]$
and thus, in general, they vary among species.
These choices are motivated by earlier studies that 
found $p^{3/2}$ dependence for a mixture of massless quarks and gluons with 
small-angle $1\leftrightarrow 2$ interactions\cite{Dusling_deltaf},
and also close to $p^{3/2}$ dependence for single-component and 
two-component systems of massless particles with energy-independent, isotropic
$2\to 2$ cross sections\cite{dyngrad}.
The two new forms here have weaker momentum dependence than 
the quadratic Grad correction,
therefore at high $p_T$ they will in general exhibit smaller dissipative
effects than the dynamical Grad results. For example, elliptic flow
is less suppressed at high $p_T$.

\begin{figure}[h]
\includegraphics[width=0.49\linewidth]{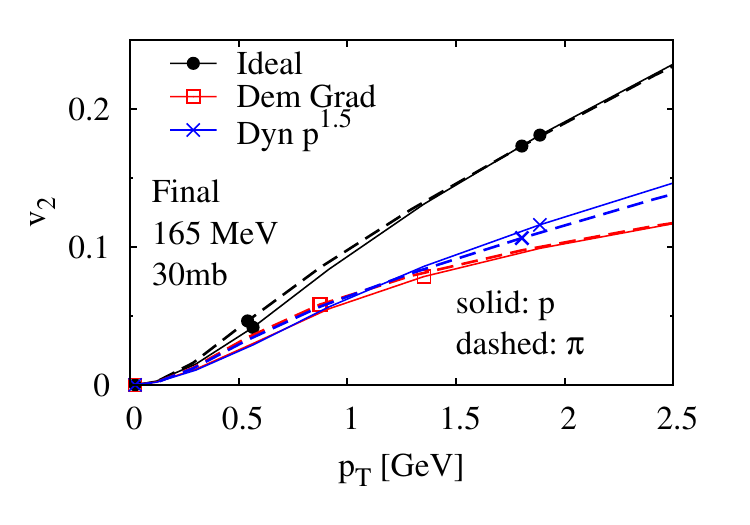}
\includegraphics[width=0.49\linewidth]{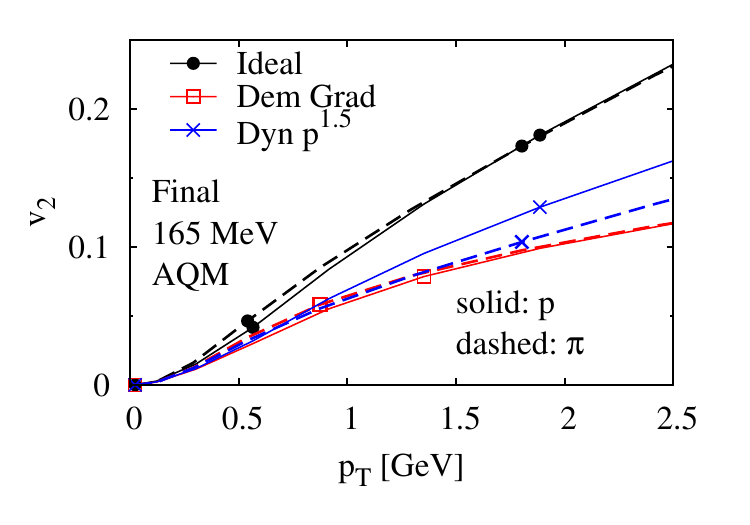}
\caption{Differential elliptic flow $v_2(p_T)$ 
for pions and protons in $Au+Au$ at 
$\sqrt{s_{NN}}=200$~GeV at RHIC with impact parameter $b=7$~fm, using
2+1D boost invariant hydrodynamic solutions from 
AZHYDRO\cite{AZHYDROrefs,AZHYDRO}, followed by Cooper-Frye fluid-to-particle
conversion
at $T_{conv} = 165$~MeV
with either the standard ``democratic'' approach (open boxes) 
or self-consistent shear corrections (crosses) with momentum
dependence $\delta f \propto p^{3/2}$
computed from kinetic theory for a gas of hadrons up to $m=1.672$~GeV.
In both plots, dashed lines are for
pions, while solid curves are for protons, and feeddown from resonance decays 
is included. 
{\em Left plot:} scenario with constant 30-mb hadronic cross sections.
{\em Right plot:} cross sections based on the additive quark model (AQM).
Results with uncorrected, local equilibrium
phase space distributions ($\delta f = 0$) are also shown (filled
circles). 
}
\label{Fig:HG165_v2pt_decays_p15}
\end{figure}

\begin{figure}[h]
\includegraphics[width=0.49\linewidth]{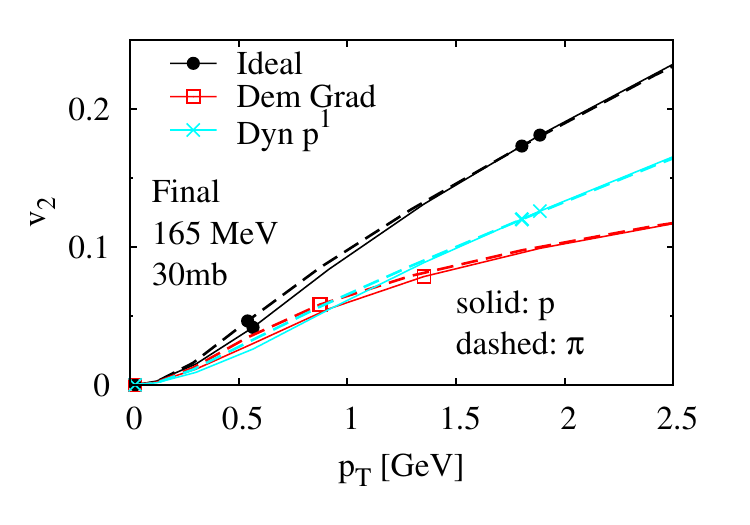}
\includegraphics[width=0.49\linewidth]{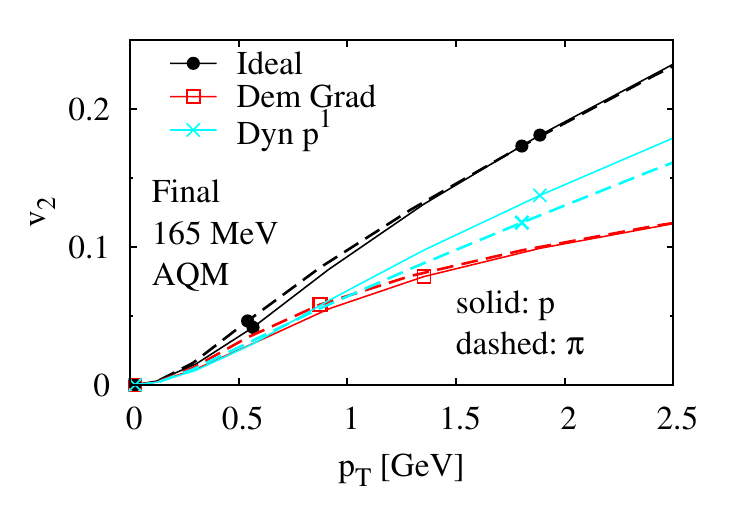}
\caption{Same as Fig.~\ref{Fig:HG165_v2pt_decays_p15}, except
with momentum dependence $\delta f \propto p$ for the curves 
from self-consistent
fluid-to-particle conversion (crosses).}
\label{Fig:HG165_v2pt_decays_p10}
\end{figure}

Figures~\ref{Fig:HG165_v2pt_decays_p15} and~\ref{Fig:HG165_v2pt_decays_p10} 
show pion and proton elliptic flow as a function of $p_T$ for 
the gas of hadrons up to $m=1.672$~GeV
with fluid-to-particle conversion at $T_{conv} = 165$~MeV 
using self-consistent linear $\delta f_i \propto p$, 
and $\delta f_i \propto p^{3/2}$,
respectively. For both figures, feeddown from resonance decays is included.
For the constant cross section scenario (left plots),
pions and protons have basically the same $v_2$, and
the main effect is an overall increase in $v_2$ at high $p_T$ 
by nearly 20\% and 40\% for
$p^{3/2}$ and $p^1$ momentum dependence, respectively, relative to the
common ``democratic Grad'' approach.
For the more realistic AQM scenario, we see a narrowing of
the separation between pion and proton $v_2$ as the power $n$ increases in
$\delta f \propto p^n$. At the same time, $v_2$ increases 
for both species.
With self consistent
fluid-to-particle conversion the viscous suppression
of proton elliptic flow is nearly two times smaller 
for $\delta f \propto p^{3/2}$,
and slightly more than two times smaller for $\delta f \propto p$,
relative to the democratic approach,

\begin{table}
\caption{Variational maxima of the functional $Q[\chi]$ as a function of 
temperature for a mixture of hadrons
up to $m=1.672$~GeV, with power law variational ansatz 
$\delta f_i \propto p^\alpha$, and 
zero chemical potentials, for constant cross sections $\sigma_{ij} = 30$~mb.
All values are rounded
to the two most significant digits. }
\label{Table:Q_vs_power_constant}
\begin{center}
\begin{tabular}{l c c c c}
\hline \hline
$\delta f / f^{eq}$    & T=100 & 120 & 140 & 165 MeV \cr 
\hline
$\propto p^1$ (linear) & 1.10 & 0.79 & 0.60 & 0.45 \cr
$\propto p^{3/2}$      & 1.16 & 0.83 & 0.63 & 0.47 \cr
$\propto p^2$ (Grad)   & 1.12 & 0.80 & 0.61 & 0.45 \cr
\hline \hline
\end{tabular}
\end{center}
\end{table}

\begin{table}
\caption{Variational maxima of the functional $Q[\chi]$ as a function of 
temperature for a mixture of hadrons
up to $m=1.672$~GeV, with power law variational ansatz 
$\delta f_i \propto p^\alpha$, and 
zero chemical potentials, for additive quark model\cite{AQM} (AQM) cross
sections with $\sigma_{MM} = 30$~mb (see text). All values are rounded
to the two most significant digits. }
\label{Table:Q_vs_power_AQM}
\begin{center}
\begin{tabular}{l c c c c}
\hline \hline
$\delta f / f^{eq}$    & T=100 & 120 & 140 & 165 MeV \cr 
\hline
$\propto p^1$ (linear) & 1.09 & 0.76 & 0.55 & 0.39 \cr
$\propto p^{3/2}$      & 1.15 & 0.80 & 0.58 & 0.41 \cr
$\propto p^2$ (Grad)   & 1.10 & 0.77 & 0.56 & 0.39 \cr
\hline \hline
\end{tabular}
\end{center}
\end{table}

One can check which of the three powers is most consistent, variationally, 
with the underlying microscopic dynamics by looking
at the maximum value of $Q$. 
As shown in Tables~\ref{Table:Q_vs_power_constant}-\ref{Table:Q_vs_power_AQM}, 
in the entire temperature range 
$100 < T_{conv} < 165$~MeV we studied, $p^{3/2}$ dependence is
favored
compared to both linear
and quadratic momentum dependence in $\delta f$. This should provide 
impetus for using $\delta f \propto p^{3/2}$ dependence instead of the common
quadratic ansatz in fluid dynamical calculations and hybrid models.
However, the results here underscore the need for species-dependent viscous
corrections even in that case.

\subsection{Sensitivity to shear viscosity}

\begin{figure}[h]
\includegraphics[width=0.49\linewidth]{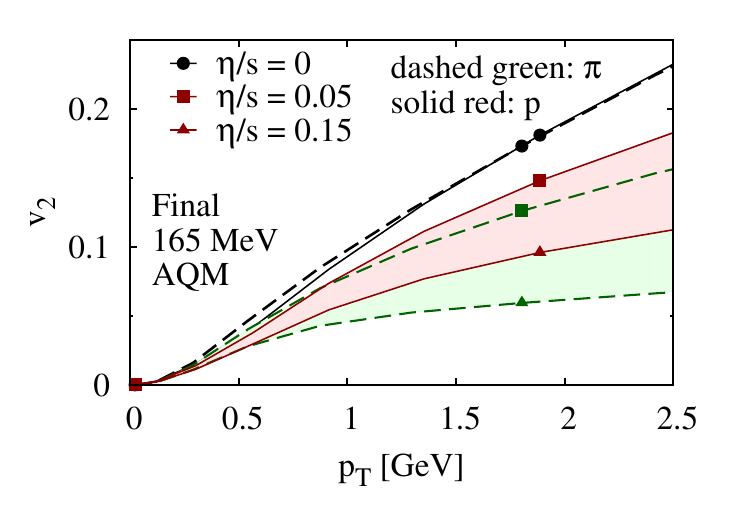}
\includegraphics[width=0.49\linewidth]{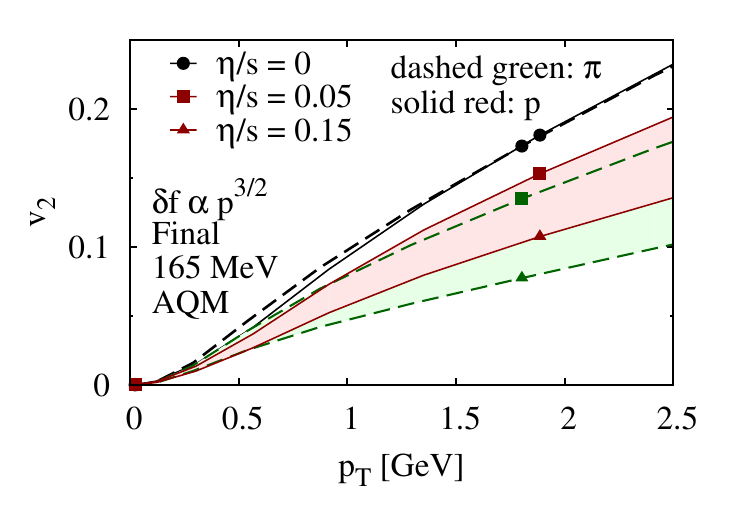}
\caption{Same as the right plots in Figs.~\ref{Fig:HG165_v2pt_decays} 
and \ref{Fig:HG165_v2pt_decays_p15} for $Au+Au$ at RHIC with self-consistent 
viscous corrections in the additive quark model (AQM) scenario
with $\delta f \propto p^2$ (left) and $\delta f\propto p^{3/2}$ (right)
but with $\eta_s/s$ varied in the range
$0.05 \le \eta_s/s \le 0.15$. Shaded bands are shown together with curves at the 
lowest $\eta_s/s = 0.05$
(squares) and highest $\eta_s/s = 0.15$ (triangles) boundaries.
}
\label{Fig:HG165_v2pt_vs_etas}
\end{figure}

The results in Secs.~\ref{Sc:species49_Grad} and \ref{Sc:species49_pto32}
correspond to a fixed set of values for hadronic cross sections,
or equivalently, a fixed shear viscosity to entropy density ratio
$\eta_s / s = 0.1$.
In Fig.~\ref{Fig:HG165_v2pt_vs_etas} we explore the sensitivity of 
differential elliptic flow $v_2(p_T)$ in $Au+Au$ at RHIC
to $\eta_s/s$ for both $p^2$ (Grad) and $p^{3/2}$
viscous corrections in the additive quark model (AQM) scenario.
All parameters are the same as in 
Figs.~\ref{Fig:HG165_v2pt_decays} and \ref{Fig:HG165_v2pt_decays_p15},
except the bands plotted for pions and protons 
correspond to $0.05 \le \eta_s/s \le 0.15$ 
($\sigma_{MM}$ is varied between 20 and 60 mb, and all hadronic cross sections
are scaled up and down proportionally).
The magnitude of 
viscous corrections in $v_2(p_T)$ 
relative to the ideal (nonviscous) case, of course,
varies with $\eta_s/s$.
In fact, the dependence on $\eta_s/s$ is 
monotonic, with the top of the bands always corresponding 
to the lowest value $\eta_s/s = 0.05$.
Still, the pion-proton splitting in $v_2(p_T)$ 
due to the self-consistent viscous corrections
is present at all $\eta_s/s$ values, and the relative difference between
pion and proton viscous corrections stays roughly the same.

\section{Conclusions}
\label{Sc:concl}

Reliable extraction of medium properties from heavy-ion data
using hydrodynamics or hybrid hydrodynamics+transport models
inevitably requires conversion of a dissipative fluid to particles (hadrons).
The popular approach is to apply the Cooper-Frye formula (\ref{CooperFrye})
with hadron phase space densities $f_i = f_i^{eq} + \delta f_i$ 
that include nonequilibrium corrections of quadratic form in momentum 
with a universal species independent 
coefficient (``democratic Grad'' ansatz). This simple scheme ignores
the dynamics of equilibration in the hadron gas. In this work we obtain
instead self-consistent shear viscous corrections 
from linearized kinetic theory (Section~\ref{Sc:framework}). 
This approach in general
gives species-dependent phase space corrections $\delta f_i$,
which are then reflected in identified particle observables. The
effect on identified particle elliptic 
flow is demonstrated in Section~\ref{Sc:multi}.

Phenomenological applications are necessarily numerical because of
the many species involved. But to aid
with interpretation we discuss extensively analytic and numerical results 
for massless and massive
two-component systems in Sections~\ref{Sc:massless} and \ref{Sc:massive_2comp}.
We also provide a comparison to fully nonlinear covariant transport to
justify the approach.

Several simplifications are made in this work, which will be 
improved in future publications. For example, realistic
energy-dependent hadronic cross sections 
and realistic viscous hydrodynamic evolution will, of course,
have to be included.
The momentum dependence of viscous corrections $\delta f_i / f^{eq}_i$ 
is also simplified here to quadratic or power-law form in momentum.
Nevertheless, it would be very interesting to check how the self-consistent
viscous distributions obtained here influence observables from 
hydrodynamic and hybrid models, and the interpretation of heavy-ion data.
To aid this we provide scaling factors in
Appendix~\ref{App:tables} that can be used to ``patch'' the commonly
used democratic approach with the species-dependent viscous corrections
calculated in this work.

\acknowledgments Insightful discussions with Gabriel Denicol,
Derek Teaney, Guy Moore, Sangyong Jeon, and Raju Venugopalan are acknowledged.
D.M. thanks RIKEN, Brookhaven National Laboratory, and the US Department of 
Energy for providing facilities essential for the completion of this work.
D.M. also thanks the hospitality of the 
Wigner Research Center for Physics (Budapest, Hungary), 
and the Institute for Nuclear Theory (Seattle, Washington), 
where parts of this work have been done. 
Computing
resources managed by RCAC/Purdue are also gratefully acknowledged.
This work was supported by the U.S. 
Department of Energy, Office of Science, under grants DE-AC02-98CH10886 
[RIKEN BNL] and \uppercase{de-sc}0004035.

\appendix

\section{General form of {\boldmath $\phi$}}
\label{App:solveChi}

The form (\ref{chi_def}) comes from expanding $\phi_i(x,\vp)$ 
in terms of irreducible tensors\cite{irred_tensors}
\be
\phi_i(x,\vp) = \sum\limits_{r = 0}^\infty 
a_r(|\tilde \vp|)\, P^{(r)}(p) \cdot X^{(r)}(x) \ ,
\label{phi_Pl}
\ee
which is just a Lorentz covariant way to write an expansion over spherical 
harmonics in the LR frame 
(the $(\cdot)$ denotes full contraction of tensors 
$P^{(r)}$ and $X^{(r)}$).
 $P^{(r)}$ is a rank-$r$ irreducible tensor projected out from
the fully symmetric, rank-$r$ 
Lorentz tensor $p^{\mu_1} p^{\mu_2} \cdots p^{\mu_r}$ such that 
$P^{(r)}$ is purely spatial in the LR frame (orthogonal to $u$ in any index)
and vanishes under contraction of any two of its indices,
so it is the irreducible representation with maximal angular momentum $r$
from the tensor product of $r$ three-dimensional (spin-1) vectors in the
LR frame,
$\overbrace{\tilde\vp \otimes \tilde\vp \otimes  \cdots \otimes \tilde\vp}^r$.
For example, with suitable normalization, $P^{(2)}_{\mu\nu}(p) = P_{\mu\nu}$
defined in (\ref{def_P_X}).
Because $\phi_i$ is a Lorentz scalar,
$X^{(r)}$ is also a rank-$r$ irreducible tensor, while the coefficients
$a_r$ are invariant under rotations in the LR frame, so their momentum
dependence is only through the LR-frame particle energy, or equivalently,
the normalized momentum magnitude $|\tilde\vp|$.
The expansion (\ref{phi_Pl}) can be inverted for $X^{(r)}$
through integration using the orthogonality of invariant tensors:
\be
X^{(r)}(x) \propto \int \frac{d^3 p}{E} P^{(r)}(p) \phi_i(x,\vp) \ ,
\ee
where the
omitted proportionality constant depends on $|\tilde\vp|$.
Inverting both sides of (\ref{linBTE_source}), the
shear source term (\ref{source_shear}) only contributes for
$r = 2$, and the result is proportional to $X^{\mu\nu}$, so
the RHS must give a similar contribution only for $r =2$.
Because the linearized collision operator commutes with
Lorentz transformations, contains scalar functions of momentum,
and $\feq_i$ only depends on $|\tilde\vp|$,
the collision operator preserves the expansion (\ref{phi_Pl}) except 
for the coefficients $a_r$. Thus, (\ref{chi_def}) indeed follows.

\section{Calculation of momentum integrals in {\boldmath $Q[\chi]$}}
\label{App:Qintegrals}

All required integrals are scalars, so it is convenient to integrate 
momenta $3$ and $4$ in the center-of-mass (CM) of the scattering process
(momentum conservation is simpler), while momenta $1$ and $2$ in the LR
frame of the fluid (so that $\feq \propto e^{-E/T}$ is
isotropic).
For brevity, in this entire Section  
$LR$ subscripts are omitted, while
$CM$ variables are distinguished with an overbar wherever 
confusion might arise.
Spherical coordinates are also helpful. 

$B_1$ can be reduced to one dimensional integration,
$Q_{11}$ and $Q_{22}$ to three dimensions, while
$Q_{31}$ and $Q_{41}$ to five dimensions in general,
or four in the case of isotropic cross sections.
All remaining integrals were performed numerically 
using adaptive integration routines
from the GNU Scientific Library (GSL) \cite{GSL}.

\subsection{\boldmath Reduction of terms $B$, $Q_{11}$, and $Q_{21}$}

The source term $B_i$ in (\ref{Qdef}), which is linear in $\chi_i$,
immediately reduces this way to
\be
B_i = \frac{2\pi}{3T^6} \int\limits_{m_i}^\infty 
       dE_1 \, p_1^5 \feq_{1i} \chi_{1i}
 \ .
\ee

In the terms quadratic in $\chi$, $\bar \vp_4$ can be eliminated using 
the $\delta$-function in three-momentum, and the magnitude of $|\bar \vp_3|$
is set by the $\delta$-function in energy:
\be
\int\limits_3\!\!\!\! \int\limits_4 \!\!
 \delta^4(12-34)\ (...)
= \frac{1}{4} \int d\bar\Omega_3 d\bar p_3
  \frac{\bar p_3^2}{\bar E_3 \bar E_4} 
  \delta(\bar E_3 + \bar E_4- \sqrt{s})\  (...)
= \left.\frac{p'_{cm}}{4\sqrt{s}} 
\, \int d\bar \Omega_3 \ (...) \right|_{\bar p_3 = p'_{cm}} \ .
\ee
For the $\chi_{1i}^2$ and $\chi_{1i}\, \chi_{2j}$ terms one
can substitute
(\ref{W_with_sigma}) to obtain
\be
\int\limits_{34} \delta^4(12-34) \bar W_{12\to 34}^{ij\to k\myell} =
p_{cm} \sqrt{s}\, (1 + \delta_{k\myell})\,\sigma_{TOT}^{ij\to k\myell}(s)  \ ,
\label{int34_W}
\ee
and the calculation is then analogous to the scattering rate
in Appendix~\ref{App:rates}. Keeping 
$t_{12} \equiv \cos\theta_{12}$, one has 
\be
Q_{11}^{ij\to k\myell} = \frac{2\pi^2}{3T^8} (1+\delta_{k\myell})
\int\limits_{m_i}^\infty dE_1\, p_1^5\, \feq_{1i}\,
\chi_{1i}^2
\int\limits_{m_j}^\infty dE_2\,  p_2\, \feq_{2j}
  \int\limits_{-1}^{1} {dt_{12}} \, F(s)\, \sigma_{TOT}^{ij\to k\myell}(s) 
\label{Q11_integration}
\ee
and
\be
Q_{21}^{ij\to k\myell} = \frac{\pi^2}{3T^8}
 (1+\delta_{k\myell})
\int\limits_{m_i}^\infty dE_1\, p_1^3\,\feq_{1i}\,
\chi_{1i}
\int\limits_{m_j}^\infty dE_2\, p_2^3\,\feq_{2j}\,
\chi_{2j}
  \int\limits_{-1}^1 dt_{12} \, (3t_{12}^2 - 1)\,
 F(s)\, \sigma_{TOT}^{ij\to k\myell}(s) \ ,
\label{Q21_integration}
\ee
where $F$ is given by (\ref{Fs}).

\subsection{\boldmath Reduction of terms $Q_{31}$ and $Q_{41}$}

The last two $\chi_1 \chi_3$ and $\chi_1 \chi_4$ terms in general
involve numerical integration in 9-4=5 dimensions
(three three-dimensional momentum integrals with a four-dimensional $\delta$-function constraint)
because $\chi_3$ and $\chi_4$ depend on outgoing 
three-momenta in the LR frame. Interchange symmetry (\ref{W_symmetry})
with $3\leftrightarrow 4$, $k\leftrightarrow \myell$ implies
$Q_{41}^{ij\to k\myell} = 
Q_{31}^{ij\to \myell k}$,
so it is enough to discuss $Q_{31}$.
For isotropic cross section, it is possible to do one more integral 
analytically,
if the LR frame momentum $\vp_3$ is expressed using
the CM frame momentum $\bar\vp_3 \equiv p'_{cm} \bar\vn_3$ 
(here $|\bar\vn_3| = 1$). Lorentz boost from CM to LR gives
\be
E_3 = \gamma_3 E_T + \beta_3 \vp_T \bar\vn_3 \ , \qquad
\vp_3 = p'_{cm} \bar\vn_3
  + \vp_T
    \left(\gamma_3 + \beta_3 \,
                        \frac{\vp_T \bar\vn_3}{E_T + \sqrt{s}}
                        \right) \ ,
\ee
where
\be
\beta_3 \equiv \frac{p'_{cm}}{\sqrt{s}} \quad, \qquad
\gamma_3 \equiv \frac{\bar E_3}{\sqrt{s}} 
= \sqrt{\beta_3^2 + \frac{m_k^2}{s}} \quad, \qquad
E_T \equiv E_1 + E_2 \quad , \qquad 
\vp_T \equiv \vp_1 + \vp_2
\ee
only depend on $\vp_1$ and $\vp_2$ but not on $\bar\vp_3$.
With convenient angles 
$\bar\vn_3(\phi_3,\theta_3)$
for the $d\bar\Omega_3$ integration such that the zenith direction is 
parallel to $\vp_T$,
\be
\bar \vn_3 \vp_T = p_T \cos\theta_3 \ ,
\qquad
\bar \vn_3 \vp_1 = p_1 (\sin\theta_1 \sin\theta_3 
\cos\phi_3 + \cos\theta_1 \cos\theta_3) \ ,
\ee 
where 
\be
\cos \theta_1 \equiv \frac{\vp_T \vp_1}{p_T p_1} 
= \frac{p_1 + p_2 t_{12}}{p_T} \ .
\ee
Because $|\bar \vp_3|$ does not depend on $\phi_3$,
the only $\phi_3$ dependence is in the $(\vp_3 \vp_1)^2$ term
from $P_3\cdot P_1$, which can be integrated.
So even if the total cross section depends on energy,
we have only four integrals remaining:
\be
\int\limits_1\!\!\!\!
\int\limits_2\!\! \int d\bar \Omega_3 \,(...)
= \frac{4\pi \cdot 2\pi \cdot 2\pi}{4} 
\int\limits_{m_i}^\infty dE_1\, p_1 
\int\limits_{m_j}^\infty dE_2\, p_2
\int\limits_{-1}^1 dt_{12} 
\int\limits_{-1}^1 dt_3 \, \langle(...)\rangle_{\phi_3}
\ee
i.e.,
\be
Q_{31}^{ij\to k\myell} = \frac{\pi^2}{2T^8}
(1+\delta_{k\myell})
\int\limits_{m_i}^\infty dE_1\, p_1\,\feq_{1i}\, \chi_{1i}
\int\limits_{m_j}^\infty dE_2\, p_2\,\feq_{2j}
\int\limits_{-1}^1 dt_{12} F(s) \sigma_{TOT}^{ij\to k\myell}(s) 
\int\limits_{-1}^1 dt_3 \, \chi_{3k}\,
\langle P_3 \cdot P_1\rangle_{\phi_3} \ ,
\label{Q31_integration}
\ee
where $t_3 \equiv \cos\theta_3$, 
$p_3 = |\vp_3| = \sqrt{(\gamma_3 E_T
      + \beta_3 p_T t_3)^2 - m_k^2}$, and 
\be
\langle (...) \rangle_{\phi_3} \equiv \frac{1}{2\pi} 
\int\limits_0^{2\pi} d\phi_3 (...)
\ee
denotes averaging over $\phi_3$. The following $\phi_3$ averages appear:
\be
\langle \bar\vn_3 \vp_1 \rangle_{\phi_3} =
p_1 \cos\theta_1 t_3 \quad , \qquad
\langle (\bar\vn_3 \vp_1)^2 \rangle_{\phi_3} =
\frac{p_1^2}{2}
[(3t_3^2 - 1) \cos^2 \!\theta_1 + 1 -  t_3^2] \ ,
\ee
in terms of which
\bea
\langle P_3\cdot P_1 \rangle_{\phi_3} &=&\frac{1}{T^4}\left[
(p'_{cm})^2 \langle (\bar\vn_3 \vp_1)^2 \rangle_\phi
+ p_1^2 (p_1 + p_2 t_{12})^2
\left(\gamma_3 + \beta_3\,
                        \frac{p_T t_3}{E_1 + E_2 + \sqrt{s}}
                        \right)^2 \right.
\nonumber \\
&&\ \qquad
\left. +\ 2p_{cm}'  p_1 (p_1 + p_2 t_{12}) 
    \left(\gamma_3 + \beta_3\,
                        \frac{p_T t_3}{E_1 + E_2 + \sqrt{s}}
    \right)
   \langle \bar\vn_3 \vp_1 \rangle_\phi \right] - \frac{p_1^2 p_3^2}{3T^4}
\label{P3P1_phi3_av}
\eea

\subsection{Integration using auxiliary variable $\omega$}

The method outlined above is practical but limited to isotropic cross section.
For general $d\sigma(s,t)/dt$,
one can evaluate $Q_{31}$ and $Q_{41}$ via extending 
the technique used in Ref.~\cite{AMYtrcoeffs} to massive particles. 
The key elements of that technique are splitting
the energy conservation integral with the help of the energy transfer $\omega$ as
\be
\delta(E_1 + E_2 - E_3 - E_4) \equiv 
\int\limits_{-\infty}^{\infty} \delta(\omega + E_1 - E_3) \, 
\delta(\omega - E_2 + E_4) \ , 
\ee
eliminating $\vp_4$ through momentum conservation, and swapping $\vp_3$ for
the momentum transfer
$\vq \equiv \vp_3 - \vp_1$.  
Exploiting rotation
invariance, introduce angles such that
\be
\vq = q(0,0,1) \ , \quad
\vp_1 = p_1 (\sin \theta_{1q}, 0, \cos\theta_{1q})\ , \quad
\vp_2 = p_2 (\cos\phi \sin \theta_{2q}, \sin\phi \sin\theta_{2q}, 
\cos\theta_{2q})  \ .
\ee
Then the Mandelstam variables for the scattering process are
\be
s = m_i^2  + m_j^2 + 2 (E_1 E_2 - \vp_1 \vp_2) \quad , \qquad 
t = \omega^2 - q^2 \ ,
\ee
the magnitudes of outgoing momenta are
\be
p_3 = \sqrt{(E_1+\omega)^2 - m_k^2} \ , \quad
p_4 = \sqrt{(E_2-\omega)^2 - m_\myell^2} \ ,
\ee
and the scalar products that appear in $s$ and $P\cdot P$ are 
\bea
\vp_1 \vp_2 &=& p_1 p_2 (\cos\theta_{1q}\cos\theta_{2q}
                      + \cos\phi \sin\theta_{1q}\sin\theta_{2q}), \nonumber \\
\vp_1 \vp_3 &=& p_1^2 + \frac{m_i^2 - m_k^2 + 2E_1 \omega + t}{2}, \ \nonumber \\
\vp_1 \vp_4 &=& p_1^2 + \vp_1 \vp_2 - \vp_1 \vp_3 \ ,
\eea
where the $\theta$ angles are fixed by the $\delta$-functions:
\be
\cos\theta_{1q} = \frac{m_i^2 - m_k^2 + 2 E_1 \omega + t}{2 p_1 q}
\ , \qquad 
\cos\theta_{2q} = \frac{m_\myell^2 - m_j^2 + 2E_2\omega - t}{2p_2 q} \ .
\ee
Five integrals remain:
\be
\int\limits_1\!\!\!\! \int\limits_2\!\!\!\! 
\int\limits_3\!\!\!\! \int\limits_4\!\!
\delta^4(12-34) \ (...)
= \frac{\pi^2}{2} 
\int\limits_{m_i}^\infty dE_1
\int\limits_{m_j}^\infty dE_2
\int\limits_0^{2\pi} d\phi
\int\limits_0^\infty dq 
\int\limits_{-\infty}^{\infty} d\omega
\,\Theta(1 - \cos^2 \!\theta_{1q}) \,\Theta(1 - \cos^2\!\theta_{2q}) (...)\ ,
\ee
where the Heaviside functions set the integration limits.

For equal masses $m_i = m_j = m_k = m_\myell \equiv m$,
\be
\int\limits_1\!\!\!\! \int\limits_2\!\!\!\! 
\int\limits_3\!\!\!\! \int\limits_4\!\!
\delta^4(12-34) \ (...)
= \frac{\pi^2}{2}
\int\limits_0^\infty dq \int\limits_{-q}^q d \omega
\int\limits_{\bar\Lambda(q,-\omega)}^\infty dE_1
\int\limits_{\bar\Lambda(q,\omega)}^\infty dE_2
\int\limits_0^{2\pi} d\phi \ (...) \ ,
\ee
where 
\be
\bar\Lambda(q,\omega) = \sqrt{m^2+ \Lambda^2(q,\omega)} \quad,\qquad
\Lambda(q,\omega) = \left| \frac{q+\omega\sqrt{1-\frac{4m^2}{t}}}{2} \right| \ ,
\ee
and we verified that both methods give numerically identical
results with isotropic cross sections. The main disadvantage
compared to the method in the previous Subsection
is speed - for isotropic cross section one still has five numerical integrals 
to do compared to four in (\ref{Q31_integration}).

\section{Evaluation of scattering rates}
\label{App:rates}

The scattering rate integral (\ref{tauij_inv}) 
right away reduces from six dimensions to only three because
in the static case the phase space density $\feq \propto e^{-E/T}$ and
Mandelstam 
\be
s \equiv m_i^2 + m_j^2 + 2 (E_1 E_2 - \vp_1 \vp_2) 
\ee
only depend on the magnitudes of momenta and the 
angle $\theta_{12}$ between them. Replacing $\cos \theta_{12}$ with $s$, 
in spherical coordinates we then have
\be
\int \frac{d^3 p_1}{E_1} \frac{d^3 p_2}{E_2} (...)
  = 4\pi \cdot 2\pi \int\limits_{m_i}^\infty dE_1 
                    \int\limits_{m_j}^\infty dE_2
                    \int\limits_{s_-}^{s_+} ds\, (...)
\ee
with limits $s_\pm = m_i^2 + m_j^2
        + 2\left[E_1 E_2 \pm \sqrt{(E_1^2 - m_i^2)(E_2^2 - m_j^2)}\right]$.

Though not pursued here, 
further simplification of the $2\to 2$ scattering rate is possible.
If speed of evaluation is a concern, consult Appendix A of 
Ref.~\cite{Pang_rate} (integrated rate for equal mass particles), 
Appendix B of Ref.~\cite{Tomasik_rate} (rate for fixed particle momentum), 
or Ref.~\cite{Huovinen_rate} (integrated rate for arbitrary masses).

\section{Grad results in nonrelativistic limit}
\label{App:Grad_NR}

In the nonrelativistic limit one can replace terms 
in (\ref{Q11_integration}),
(\ref{Q21_integration}), and (\ref{Q31_integration}) with their
nonrelativistic counterparts
\be
\frac{d^3p}{E} \to \frac{d^3p}{m} \ , \quad 
dE \to \frac{dp\, p}{m} \ , \quad
\exp\!\left(-\frac{E}{T}\right) 
 \to \exp\!\left(-\frac{m}{T} - \frac{p^2}{2mT}\right) \ , \quad
F(s) \to m_i m_j |\vv_1 - \vv_2| \ .
\ee
Similarly, in (\ref{P3P1_phi3_av})
\be
\gamma_3 \to \frac{m_3}{m_1+m_2} = \frac{m_3}{m_3+m_4} \ , \qquad 
 \beta_3 \to 0 \ , \qquad
p_3^2 \to (p_{cm}')^2 + \gamma_3^2 p_T^2 + 2p_{cm}' \gamma_3 p_T t_3 \ .
\ee
Note that it is simpler to get the above result for $p_3$ from 
$\vp_3 \approx \bar \vp_3 + \gamma_3 \vp_T$ than from 
$\sqrt{E_3^2 - m_k^2}$ because 
there is an almost perfect cancellation in the latter.

It is further convenient to switch variables from $\vp_1$ and $\vp_2$
to total momentum and relative
velocity
\be
\vp_T = \vp_1 + \vp_2\ , \  
\vv_{rel} \equiv \vv_1 - \vv_2  \qquad \Leftrightarrow \qquad
\vp_1 = \frac{m_1}{m_1 + m_2} (\vp_T + m_2 \vv_{rel}) \ , \
\vp_2 = \frac{m_2}{m_1 + m_2} (\vp_T - m_1 \vv_{rel}) \ , \
\ee
for which
\bea
d^3 p_1 d^3 p_2 = \left(\frac{m_1 m_2}{m_1 + m_2}\right)^3 d^3 p_T d^3 v_{rel}
\eea
so
\be
\int\limits_{m_i}^\infty dE_1 \int\limits_{m_j}^\infty dE_2 
\int\limits_{-1}^1 dt_{12}\, F(s)\, (...)
\quad \to \quad \left(\frac{m_i m_j}{m_i+m_j}\right)^3
                 \int\limits_0^\infty dp_T \int\limits_0^\infty dv_{rel} 
                 \int\limits_{-1}^1 d\cos\tilde\theta\  
                   \frac{p_T^2 v_{rel}^3}{p_1 p_2}\ (...) \ ,
\ee 
where $\tilde\theta$ is the angle between $\vp_T$ and $\vv_{rel}$,
while in the exponents
\be
\frac{p_1^2}{2 m_1 T} + \frac{p_2^2}{2 m_2 T} 
  = \frac{p_T^2 + m_1 m_2 v_{rel}^2}{2 (m_1 + m_2) T}  \ .
\ee
Straightforward integration leads then to (\ref{Qvalues_NR}).

\section{Longitudinal boost invariance and Cooper-Frye integrals}
\label{App:CooperFrye}

For longitudinally boost invariant systems (cf. footnote [44])
hyperbolic $\eta \equiv \frac{1}{2} \ln \frac{t + z}{t - z}$ 
and $\tau \equiv \sqrt{t^2 - z^2}$ coordinates are most convenient 
for spacetime,
while rapidity $y \equiv \frac{1}{2} \ln \frac{E + p_z}{E - p_z}$
and transverse mass $m_T \equiv \sqrt{p_T^2 + m^2}$ for momenta:
\be
x^\mu = (\tau\, \ch\, \eta, \vx_T, \tau\, \sh\, \eta) \quad , \qquad
p^\mu = (m_T\, \ch\, y, \vp_T, m_T\, \sh\, y) \ .
\label{def_x_p}
\ee
The Cooper-Frye formula for the distribution of particles 
emitted from a surface element
$d\sigma^\mu$ of a 3D spacetime hypersurface is
\be
E\frac{dN_i(x,\vp)}{d^3p} \equiv \frac{dN_i(x,\vp_T,y)}{d^2 p_T dy} 
  = p^\mu d\sigma_\mu(x)  f_i(x,\vp) \ .
\label{CooperFrye}
\ee
Often a $\Theta(p^\mu d\sigma_\mu)$ factor is also 
included
to cut out potential negative contributions from spacelike surface elements
but it is not used in this work. With boost invariance,
\be
d\sigma^\mu = n^\mu \tau d\eta d^2 x_T \quad , \qquad 
n^\mu = (n^0\, \ch\, \eta, \vn_T, 
             n^0\, \sh\,\eta) \ .
\ee
i.e.,
\be
p^\mu d\sigma_\mu = \tau [m_T n^0 \ch\,\xi - \vp_T \vn_T]   d\eta d^2 x_T \ ,
\ee
where $\xi \equiv \eta - y$.
In the thermal equilibrium distribution (\ref{feq})
\be
u^\mu = \gamma (\ch\, \eta, \vv_T, \sh\,\eta) \quad, \qquad 
\gamma \equiv \frac{1}{\sqrt{1-v_T^2}}
\qquad \Rightarrow \qquad (pu) = \gamma\, (m_T\, \ch\,\xi - \vp_T \vv_T) \ ,
\label{pdotu}
\ee
and in the shear correction (\ref{chi_def}),
$|\tilde\vp| = \sqrt{(pu)^2-m^2} / T$,
\be
\pi^{\mu\nu}p_\mu p_\nu = 
 m_T^2(\pi^{00} \ch^2\xi + \pi^{zz}\sh^2\xi) 
  - 2 m_T \ch\,\xi\,(p_x \pi^{0x} + p_y \pi^{0y})
  + p_x^2 \pi^{xx} + p_y^2 \pi^{yy} + 2 p_x p_y \pi^{xy} \ ,
\label{pdotpidotp}
\ee
with shear stress components all taken at $\eta=0$. For several
equivalent forms of the last expression, see Ref.~\cite{OSUvhydro}.

Boost invariant 2+1D viscous fluid dynamics provides
hydrodynamic fields ($T$, $\left\{\mu_c\right\}$, $\vv_T$, $\pi^{\mu\nu}$) 
and hypersurface elements ($n^0$, $\vn_T$) in the $\eta=0$ frame, 
as a function of $\tau$ and $\vx_T$. If one is only 
interested in the momentum distribution, one integrates
(\ref{CooperFrye})
over the hypersurface, which includes at each $\tau$ and $\vx_T$
 integration
over $\eta$:
\be
\tau \int\limits_{-\infty}^{\infty} d\eta\, 
         [m_T n^0 \ch\,\xi - \vp_T \vn_T] f_i(\tau,\vx_T,\vp_T,\xi)
= 2 \tau \int\limits_0^{\infty} d\xi\, 
     [m_T n^0 \ch\,\xi - \vp_T \vn_T] f_i(\tau,\vx_T,\vp_T,\xi)
\label{xi_integral_in_CF}
\ee
with reflection symmetry along the beam axis assumed.
For the ideal piece, (\ref{xi_integral_in_CF}) yields
\be
2 \tau \frac{g_i}{(2\pi)^3} e^{\alpha_T}
       [m_T n^0 K_1(z_T) - \vp_T \vn_T K_0(z_T)]
\quad , \qquad 
  z_T \equiv \frac{\gamma m_T}{T} 
\quad \ , \qquad 
  \alpha_T \equiv  \frac{\mu_i + \gamma \vp_T \vv_T}{T} \ .
\ee
For the viscous correction (\ref{chi_def}), the integral 
can only be evaluated analytically in special cases.
For example, for quadratic corrections in momentum (\ref{Grad}), 
one has%
\footnote{
There are many equivalent ways to write these expressions 
because of Bessel function
identities, such as $K_{n-1}(x) + 2nK_n(x) / x \equiv K_{n+1}(x)$.
}
\bea
\frac{\chi^{Grad}}{\eta_s T^3} 
  \frac{g_i}{(2 \pi)^3} 
   e^{\alpha_T} &&
   \!\!\!\!\!
   2\tau \left\{ m_T n^0
         \left[m_T^2 \left(K_1(z_T) 
               + \frac{K_2(z_T)}{z_T} \right) \pi^{00} 
        + m_T^2 \frac{K_2(z_T)}{z_T}\pi^{zz}
        - 2 m_T \left(K_0(z_T) + \frac{K_1(z_T)}{z_T} \right) 
                \left(p_x \pi^{0x}+p_y \pi^{0y} \right)
         \right. 
   \right.
\nonumber \\
&& \qquad \qquad + \left.
         \left. 
       K_1(z_T) 
          \left(p_x^2 \pi^{xx}+p_y^2 \pi^{yy}+2 p_x p_y \pi^{xy} \right) 
        \bigg] \right.
    \right.
\nonumber \\
&&  - \left.
           \vp_T \cdot \vn_T 
         \left[m_T^2 \left(K_0(z_T)
                           +\frac{K_1(z_T)}{z_T} \right)\pi^{00}
               + m_T^2 \frac{K_1(z_T)}{z_T} \pi^{zz}
               - 2 m_T K_1(z_T) \left(p_x \pi^{0x}+p_y \pi^{0y} \right) 
         \right.
      \right.
\nonumber \\
&& \qquad \qquad \qquad + \left. \left. K_0(z_T) 
     \left(p_x^2 \pi^{xx}+p_y^2 \pi^{yy}+2 p_x p_y \pi^{xy} \right) 
      \bigg] \right. \bigg\} \right.
\eea

\section{Flow anisotropies for viscous 
four-source model}
\label{App:foursource}

Here we calculate differential harmonic flow coefficients $v_n(p_T)$ 
for the four-source model of Sec.~\ref{Sc:foursource}.
For isochronous emission from a spatially uniform fireball, 
the momentum distribution of particles is $dN/d^3 p = V f$. By assumption, 
the laboratory-frame volume $V$ is the same for all four
fireballs in the model. Below we construct $f$ using viscous
corrections $\phi$ of the Grad form (\ref{phi_dem}) for each source, 
and then evaluate $v_n(p_T)$ via (\ref{def_vnpt}). It is sufficient to calculate
$f_{(+x)}$ because $f$ for the other three sources can be obtained via 
suitable rotation and/or mirror 
symmetry in the transverse plane. Unless noted otherwise, all 
vectors and tensors below are in laboratory (observational) frame coordinates.

For the source moving with a three-velocity $(v_x, 0, 0)$,
the equilibrium phase space distribution 
evaluated at a general on-shell four-momentum $p$  with azimuth such that
$\vp_T \equiv p_T(\cos\phi, \sin\phi)$ is
\be
f^{\rm eq} = N e^{-\gamma_x m_T \ch\, y/T} e^{\gamma_x v_x p_T \cos \phi/T} \ ,
\qquad
N\equiv\frac{g}{(2\pi)^3} e^{\mu/T} \ ,
\qquad \gamma_x \equiv \frac{1}{\sqrt{1-v_x^2}}
\ee
(cf. (\ref{feq}), (\ref{def_x_p}), and (\ref{pdotu})).
The normalization $N$ and the volume $V$ drop out in the anisotropy
coefficients.
The Grad viscous corrections depend
on $\pi^{\mu\nu}p_\mu p_\nu$ given by
(\ref{pdotpidotp}). Instead of boosting (\ref{pimunu_LR_Bjorken}) 
to obtain the shear stress tensor
for the fireball, it is simpler to evaluate this scalar via inverse
boosting $p$ to the fluid rest frame, where
\be
p^\mu_{LR} = (\gamma_x (m_T \ch\, y  -v_x p_T \cos\phi), 
              \gamma_x (p_T\cos\phi - v_x m_T\ch\, y), p_T\sin\phi, m_T \sh\, y) \ ,
\ee
and $\pi^{\mu\nu}_{LR}$ is diagonal.
Straightforward algebra then yields, at midrapidity $y = 0$,
\be
f_{(+x)} 
 = N e^{-a_x} e^{b_x \cos \phi} 
       \left[1 
        + \frac{c \kappa}{4} 
          (z^2 - a_x^2 + 2 a_x b_x \cos\phi - b_x^2 \cos^2\phi)  
       \right]
\ee
with
\be
\kappa \equiv \frac{\pi_L}{e+p} \ ,
\qquad
z \equiv \frac{m}{T} \ ,
\ee
and shorthands
\be
a_x \equiv \frac{\gamma_x m_T}{T} \ , \qquad 
b_x \equiv \frac{\gamma_x v_x p_T}{T} \ .
\ee
Species dependence enters through the mass in $z$ and $a_x$, 
and for dynamical Grad corrections also through $c$. For 
the democratic Grad ansatz, only the mass matters because $c = 1$ 
is set for all species.

Ninety-degree rotation $\phi \to \phi - \pi/2$ and substitution $v_x\to v_y$
gives $f_{(+y)}$, and similar rotations by $\pi$ and $3\pi/2$, or equivalently,
reflections $v_x\to -v_x$, $v_y\to -v_y$, give the remaining two source
distributions. Thus, at midrapidity,
\be
f(p_T,\phi) = f_{(+x)} + \left.f_{(+x)}\right|_{b_x\to -b_x}
              + \left.f_{(+x)}\right|_{\substack{
                                       a_x\to a_y\hfill \\
                                       b_x\to b_y\hfill \\
                                       \cos\phi \to \sin\phi}}
              + \left.f_{(+x)}\right|_{\substack{
                                       a_x\to a_y\hfill \\
                                       b_x\to -b_y\hfill \\
                                       \cos\phi \to \sin\phi}} \ .
\ee 
Harmonic flow coefficients (\ref{def_vnpt}) can now be readily evaluated.
Each term in the denominator reduces to the integral over $f_{(+x)}$ via
shifts $\phi \to \phi - \pi/2$, $\phi \to \phi - \pi$, $\phi \to \phi - 3\pi/2$ in the integrals for $f_{(+y)}$, $f_{(-x)}$, $f_{(-y)}$, respectively,
which do not affect 
the range of integration. Analogous shifts of $\phi$ in the numerator 
have the potential side effect of changing the sign of $\cos(n\phi)$.
For odd $n$, a shift by $\pi$ brings a minus sign and,
therefore, contributions
cancel, i.e., $v_n = 0$ 
(this is also evident from the symmetry of the configuration).
For even $n$, a shift by $\pi$ preserves the sign, so the sources moving
along the $\pm x$ directions contribute equally. Shifts by 
$\pi/2$ and $3\pi/2$ flip sign in the numerator
whenever $n$ is not divisible by 4,
so the sources moving along the $\pm y$ 
direction also contribute equally but with potentially opposite overall sign.
Therefore, for even $n$,
\be
v_n(p_T) = \frac{G_n(a_x,b_x,z,c\kappa) + (-1)^{n/2} G_n(a_y,b_y,z,c\kappa)
                }
                {G_0(a_x,b_x,z,c\kappa) + G_0(a_y,b_y,z,c\kappa)
                } \ ,
\label{res_vnpt}
\ee
where the shorthand
\bea
G_n(a_x, b_x, z, c\kappa)
&\equiv&
\frac{1}{2\pi N} \int\limits_0^{2\pi} d\phi \, f_{(+x)} \,\cos (n\phi)
\nonumber \\
&=& e^{-a_x} 
  \left\{I_n
      + \frac{c\kappa}{4}\left[(z^2 - a_x^2)I_n
                         + a_x b_x (I_{|n-1|} + I_{n+1}) -
                          \frac{b_x^2}{4} (I_{|n-2|} + 2 I_{n} + I_{n+2}) \right] \right\}
\label{def_G}
\eea
involves modified Bessel functions of the first kind
\be
I_n \equiv I_n (b_x) \equiv \frac{1}{2\pi}\int\limits_0^{2\pi} d\varphi\, e^{b_x \cos \varphi} \cos(n\varphi) \ .
\ee
(Integrals with $\cos \phi\, \cos(n\phi) $ and $\cos^2\phi\, \cos (n\phi)$
reduce to those with a single cosine with the help of the cosine 
addition theorem and $\cos^2\phi \equiv (\cos 2\phi + 1) / 2$.)
Note that for $\kappa \le 0$ 
the denominator in (\ref{res_vnpt}) is strictly positive because,
at midrapidity, $\pi^{\mu\nu} p_\mu p_\nu = -\pi_L (p_{x,LR}^2 + p_{y,LR}^2)/2 \ge 0$.

\section{Self-consistent Grad coefficient tables}
\label{App:tables}

Tables \ref{Table:C_Grad_constant}-\ref{Table:C_Grad_AQM}, \ref{Table:C_3over2_constant}-\ref{Table:C_3over2_AQM}, 
and \ref{Table:C_linear_constant}-\ref{Table:C_linear_AQM} tabulate 
self-consistent viscous phase space corrections
for the gas of hadrons in Section~\ref{Sc:multi},
using $\delta f/\feq \propto p^2$, $p^{3/2}$ and $p$, respectively. 
In all six tables, 
correction factors {\em relative} to the ``democratic Grad'' 
form (\ref{phi_dem}) are printed (rounded to two decimal figures). 
To apply the dynamical 
correction for species $i$,
read the coefficient $c_i$ from the table for the species 
and use Eqs. (\ref{df_corrections}) to obtain the viscous correction with 
the desired momentum dependence.

The corrections depend rather smoothly on
hadron (pole) mass, and therefore can also be well 
represented by fits of the form
\be
c(x) = \delta + \alpha \left[1 + \displaystyle{\left(\frac{x}{\gamma}\right)^\beta}\right]^{-1}
\qquad {\rm or} \qquad 
c(x) =  \alpha + \beta |x - \gamma|^\delta \ , 
\qquad x \equiv \frac{m}{1\ {\rm GeV}} \ ,
\label{Cfit_fns}
\ee 
where $x$ is the hadron (pole) mass $m$ in GeV units. 
Tables~\ref{Table:Cfits_Grad}-\ref{Table:Cfits_linear} list the 
best fit values for the parameters $\alpha$, $\beta$, $\gamma$, and $\delta$ 
as a function of temperature
for the various scenarios in
Tables \ref{Table:C_Grad_constant}-\ref{Table:C_linear_AQM}. 
The fits are done to
the original unrounded $c_i$ values. Note that there are separate fits
for mesons and baryons in the case of additive quark model (AQM) 
cross sections. There is no specific physics motivation
behind the forms (\ref{Cfit_fns}); the functions are chosen solely for accuracy
(the relative accuracy is better than $8.5\times 10^{-4}$ in all cases).

\begin{table}
\caption{Species-dependent shear viscous phase space corrections calculated
as a function of temperature
for a gas of hadrons up to $m=1.672$~GeV
with the same constant cross section for all species,
assuming quadratic momentum dependence $\delta f/f^{eq} \propto p^2$ 
(dynamical Grad approximation).}
\label{Table:C_Grad_constant}
\begin{center}
\begin{tabular}{lcccc}
\hline\hline
Species   & T = 100 & 120  & 140 & 165 MeV\cr
\hline
$\pi$ & 1.08 & 1.13 & 1.17 & 1.21 \cr
K & 0.89 & 0.96 & 1.02 & 1.08 \cr
$\eta$ & 0.87 & 0.94 & 1.00 & 1.06 \cr
$f_0$ & 0.85 & 0.92 & 0.98 & 1.04 \cr
$\rho$ & 0.80 & 0.87 & 0.93 & 0.99 \cr
$\omega$ & 0.80 & 0.86 & 0.93 & 0.99 \cr
$K^*$(892) & 0.77 & 0.83 & 0.90 & 0.96 \cr
N & 0.76 & 0.82 & 0.88 & 0.94 \cr
$\eta'$(958) & 0.75 & 0.82 & 0.88 & 0.94 \cr
$f_0$(980) & 0.75 & 0.81 & 0.87 & 0.93 \cr
$a_0$(980) & 0.75 & 0.81 & 0.87 & 0.93 \cr
$\phi$(1020) & 0.74 & 0.81 & 0.86 & 0.92 \cr
$\Lambda$ & 0.72 & 0.79 & 0.84 & 0.90 \cr
$h_1$(1170) & 0.72 & 0.78 & 0.83 & 0.89 \cr
$\Sigma$ & 0.71 & 0.77 & 0.83 & 0.89 \cr
$b_1$(1235) & 0.71 & 0.76 & 0.82 & 0.88 \cr
$\Delta$(1232) & 0.71 & 0.76 & 0.82 & 0.88 \cr
$a_1$(1260) & 0.71 & 0.77 & 0.82 & 0.88 \cr
$K_1$(1270) & 0.70 & 0.76 & 0.81 & 0.87 \cr
$f_2$(1270) & 0.70 & 0.76 & 0.81 & 0.87 \cr
$f_1$(1285) & 0.70 & 0.76 & 0.81 & 0.87 \cr
$\eta$(1295) & 0.70 & 0.75 & 0.81 & 0.87 \cr
$\pi$(1300) & 0.70 & 0.75 & 0.81 & 0.87 \cr
$\Xi$ & 0.69 & 0.75 & 0.81 & 0.86 \cr
$a_2$(1320) & 0.69 & 0.75 & 0.81 & 0.86 \cr
$\Sigma$(1385) & 0.68 & 0.74 & 0.80 & 0.85 \cr
$f_0$(1370) & 0.69 & 0.74 & 0.80 & 0.85 \cr
$K_1$(1400) & 0.68 & 0.74 & 0.79 & 0.85 \cr
$\Lambda$(1405) & 0.68 & 0.74 & 0.79 & 0.85 \cr
$K^*$(1410) & 0.68 & 0.74 & 0.79 & 0.85 \cr
$\eta$(1405) & 0.68 & 0.74 & 0.79 & 0.85 \cr
$\omega$(1420) & 0.68 & 0.74 & 0.79 & 0.84 \cr
$f_1$(1420) & 0.68 & 0.73 & 0.79 & 0.84 \cr
$K_0^*$(1430) & 0.68 & 0.73 & 0.79 & 0.84 \cr
$K_2^*$(1430) & 0.68 & 0.73 & 0.79 & 0.84 \cr
N(1440) & 0.68 & 0.73 & 0.79 & 0.84 \cr
$\rho$(1450) & 0.67 & 0.73 & 0.78 & 0.84 \cr
$f_0$(1500) & 0.67 & 0.72 & 0.78 & 0.83 \cr
$\Lambda$(1520) & 0.67 & 0.72 & 0.77 & 0.83 \cr
N(1520) & 0.67 & 0.72 & 0.77 & 0.83 \cr
$f_2'$(1525) & 0.67 & 0.72 & 0.77 & 0.83 \cr
$\Xi$(1530) & 0.67 & 0.72 & 0.77 & 0.83 \cr
N(1535) & 0.67 & 0.72 & 0.77 & 0.83 \cr
$\Delta$(1600) & 0.66 & 0.71 & 0.76 & 0.82 \cr
$\Lambda$(1600) & 0.66 & 0.71 & 0.76 & 0.82 \cr
$\Delta$(1620) & 0.66 & 0.71 & 0.76 & 0.81 \cr
$\omega$(1650) & 0.65 & 0.71 & 0.76 & 0.81 \cr
N(1650) & 0.65 & 0.71 & 0.76 & 0.81 \cr
$\Omega$ & 0.65 & 0.70 & 0.75 & 0.81 \cr
\hline \hline
\end{tabular}
\end{center}
\end{table}

\begin{table}
\caption{Species-dependent shear viscous phase space corrections calculated 
as a function of temperature
for a gas of hadrons up to $m=1.672$~GeV
with additive quark model\cite{AQM} (AQM) cross sections,
assuming quadratic momentum dependence $\delta f/f^{eq} \propto p^2$ 
(dynamical Grad approximation).}
\label{Table:C_Grad_AQM}
\begin{center}
\begin{tabular}{lcccc}
\hline\hline
Species & T = 100 & 120  & 140 & 165 MeV\cr
\hline
$\pi$ & 1.08 & 1.15 & 1.21 & 1.27 \cr
K & 0.90 & 0.98 & 1.06 & 1.14 \cr
$\eta$ & 0.88 & 0.95 & 1.03 & 1.12 \cr
$f_0$ & 0.86 & 0.94 & 1.01 & 1.10 \cr
$\rho$ & 0.80 & 0.88 & 0.96 & 1.04 \cr
$\omega$ & 0.80 & 0.88 & 0.95 & 1.04 \cr
$K^*$(892) & 0.77 & 0.85 & 0.92 & 1.01 \cr
N & 0.56 & 0.62 & 0.68 & 0.74 \cr
$\eta'$(958) & 0.76 & 0.83 & 0.91 & 0.99 \cr
$f_0$(980) & 0.75 & 0.83 & 0.90 & 0.98 \cr
$a_0$(980) & 0.75 & 0.83 & 0.90 & 0.98 \cr
$\phi$(1020) & 0.75 & 0.82 & 0.89 & 0.97 \cr
$\Lambda$ & 0.53 & 0.59 & 0.64 & 0.70 \cr
$h_1$(1170) & 0.72 & 0.79 & 0.86 & 0.94 \cr
$\Sigma$ & 0.52 & 0.58 & 0.63 & 0.69 \cr
$b_1$(1235) & 0.71 & 0.78 & 0.85 & 0.93 \cr
$\Delta$(1232) & 0.52 & 0.57 & 0.62 & 0.68 \cr
$a_1$(1260) & 0.71 & 0.78 & 0.85 & 0.93 \cr
$K_1$(1270) & 0.70 & 0.77 & 0.84 & 0.92 \cr
$f_2$(1270) & 0.70 & 0.77 & 0.84 & 0.92 \cr
$f_1$(1285) & 0.70 & 0.77 & 0.84 & 0.92 \cr
$\eta$(1295) & 0.70 & 0.77 & 0.84 & 0.91 \cr
$\pi$(1300) & 0.70 & 0.77 & 0.83 & 0.91 \cr
$\Xi$ & 0.51 & 0.56 & 0.61 & 0.67 \cr
$a_2$(1320) & 0.70 & 0.76 & 0.83 & 0.91 \cr
$\Sigma$(1385) & 0.50 & 0.55 & 0.60 & 0.66 \cr
$f_0$(1370) & 0.69 & 0.75 & 0.82 & 0.90 \cr
$K_1$(1400) & 0.68 & 0.75 & 0.82 & 0.89 \cr
$\Lambda$(1405) & 0.50 & 0.55 & 0.60 & 0.66 \cr
$K^*$(1410) & 0.68 & 0.75 & 0.82 & 0.89 \cr
$\eta$(1405) & 0.68 & 0.75 & 0.82 & 0.89 \cr
$\omega$(1420) & 0.68 & 0.75 & 0.81 & 0.89 \cr
$f_1$(1420) & 0.68 & 0.75 & 0.81 & 0.89 \cr
$K_0^*$(1430) & 0.68 & 0.75 & 0.81 & 0.89 \cr
$K_2^*$(1430) & 0.68 & 0.75 & 0.81 & 0.89 \cr
N(1440) & 0.49 & 0.54 & 0.60 & 0.65 \cr
$\rho$(1450) & 0.68 & 0.74 & 0.81 & 0.88 \cr
$f_0$(1500) & 0.67 & 0.74 & 0.80 & 0.88 \cr
$\Lambda$(1520) & 0.48 & 0.53 & 0.59 & 0.64 \cr
N(1520) & 0.48 & 0.53 & 0.59 & 0.64 \cr
$f_2'$(1525) & 0.67 & 0.73 & 0.80 & 0.87 \cr
$\Xi$(1530) & 0.48 & 0.53 & 0.58 & 0.64 \cr
N(1535) & 0.48 & 0.53 & 0.58 & 0.64 \cr
$\Delta$(1600) & 0.48 & 0.53 & 0.58 & 0.63 \cr
$\Lambda$(1600) & 0.48 & 0.53 & 0.58 & 0.63 \cr
$\Delta$(1620) & 0.48 & 0.52 & 0.57 & 0.63 \cr
$\omega$(1650) & 0.66 & 0.72 & 0.78 & 0.85 \cr
N(1650) & 0.47 & 0.52 & 0.57 & 0.63 \cr
$\Omega$ & 0.47 & 0.52 & 0.57 & 0.62 \cr
\hline \hline
\end{tabular}
\end{center}
\end{table}

\begin{table}
\caption{Species-dependent shear viscous phase space corrections calculated 
as a function of temperature
for a gas of hadrons up to $m=1.672$~GeV
with the same constant cross section for all species,
assuming power-law momentum dependence $\delta f/f^{eq} \propto p^{3/2}$.}
\label{Table:C_3over2_constant}
\begin{center}
\begin{tabular}{lcccc}
\hline\hline
Species   & T = 100 & 120  & 140 & 165 MeV\cr
\hline
$\pi$ & 2.56 & 2.68 & 2.79 & 2.87 \cr
K & 2.31 & 2.45 & 2.58 & 2.69 \cr
$\eta$ & 2.28 & 2.42 & 2.55 & 2.66 \cr
$f_0$ & 2.26 & 2.39 & 2.52 & 2.64 \cr
$\rho$ & 2.19 & 2.32 & 2.45 & 2.57 \cr
$\omega$ & 2.19 & 2.32 & 2.44 & 2.56 \cr
$K^*$(892) & 2.15 & 2.28 & 2.41 & 2.52 \cr
N & 2.14 & 2.27 & 2.39 & 2.51 \cr
$\eta'$(958) & 2.14 & 2.26 & 2.38 & 2.50 \cr
$f_0$(980) & 2.13 & 2.26 & 2.38 & 2.49 \cr
$a_0$(980) & 2.13 & 2.25 & 2.38 & 2.49 \cr
$\phi$(1020) & 2.12 & 2.24 & 2.37 & 2.48 \cr
$\Lambda$ & 2.10 & 2.22 & 2.34 & 2.45 \cr
$h_1$(1170) & 2.09 & 2.21 & 2.32 & 2.44 \cr
$\Sigma$ & 2.09 & 2.20 & 2.32 & 2.43 \cr
$b_1$(1235) & 2.08 & 2.20 & 2.31 & 2.42 \cr
$\Delta$(1232) & 2.08 & 2.20 & 2.31 & 2.42 \cr
$a_1$(1260) & 2.08 & 2.20 & 2.31 & 2.42 \cr
$K_1$(1270) & 2.07 & 2.19 & 2.30 & 2.41 \cr
$f_2$(1270) & 2.07 & 2.19 & 2.30 & 2.41 \cr
$f_1$(1285) & 2.07 & 2.19 & 2.30 & 2.41 \cr
$\eta$(1295) & 2.07 & 2.18 & 2.30 & 2.40 \cr
$\pi$(1300) & 2.07 & 2.18 & 2.29 & 2.40 \cr
$\Xi$ & 2.07 & 2.18 & 2.29 & 2.40 \cr
$a_2$(1320) & 2.07 & 2.18 & 2.29 & 2.40 \cr
$\Sigma$(1385) & 2.06 & 2.17 & 2.28 & 2.38 \cr
$f_0$(1370) & 2.06 & 2.17 & 2.28 & 2.39 \cr
$K_1$(1400) & 2.06 & 2.17 & 2.27 & 2.38 \cr
$\Lambda$(1405) & 2.06 & 2.16 & 2.27 & 2.38 \cr
$K^*$(1410) & 2.06 & 2.16 & 2.27 & 2.38 \cr
$\eta$(1405) & 2.06 & 2.16 & 2.27 & 2.38 \cr
$\omega$(1420) & 2.05 & 2.16 & 2.27 & 2.38 \cr
$f_1$(1420) & 2.05 & 2.16 & 2.27 & 2.37 \cr
$K_0^*$(1430) & 2.05 & 2.16 & 2.27 & 2.37 \cr
$K_2^*$(1430) & 2.05 & 2.16 & 2.27 & 2.37 \cr
N(1440) & 2.05 & 2.16 & 2.27 & 2.37 \cr
$\rho$(1450) & 2.05 & 2.16 & 2.26 & 2.37 \cr
$f_0$(1500) & 2.04 & 2.15 & 2.26 & 2.36 \cr
$\Lambda$(1520) & 2.04 & 2.15 & 2.25 & 2.35 \cr
N(1520) & 2.04 & 2.15 & 2.25 & 2.35 \cr
$f_2'$(1525) & 2.04 & 2.15 & 2.25 & 2.35 \cr
$\Xi$(1530) & 2.04 & 2.15 & 2.25 & 2.35 \cr
N(1535) & 2.04 & 2.15 & 2.25 & 2.35 \cr
$\Delta$(1600) & 2.03 & 2.14 & 2.24 & 2.34 \cr
$\Lambda$(1600) & 2.03 & 2.14 & 2.24 & 2.34 \cr
$\Delta$(1620) & 2.03 & 2.13 & 2.24 & 2.34 \cr
$\omega$(1650) & 2.03 & 2.13 & 2.23 & 2.33 \cr
N(1650) & 2.03 & 2.13 & 2.23 & 2.33 \cr
$\Omega$ & 2.03 & 2.13 & 2.23 & 2.33 \cr
\hline \hline
\end{tabular}
\end{center}
\end{table}

\begin{table}
\caption{Species-dependent shear viscous phase space corrections calculated 
as a function of temperature
for a gas of hadrons up to $m=1.672$~GeV
with additive quark model\cite{AQM} (AQM) cross sections,
assuming power-law momentum dependence $\delta f/f^{eq} \propto p^{3/2}$.}
\label{Table:C_3over2_AQM}
\begin{center}
\begin{tabular}{lcccc}
\hline\hline
Species & T = 100 & 120  & 140 & 165 MeV\cr
\hline
$\pi$ & 2.57 & 2.72 & 2.87 & 3.03 \cr
K & 2.32 & 2.48 & 2.66 & 2.83 \cr
$\eta$ & 2.29 & 2.45 & 2.63 & 2.81 \cr
$f_0$ & 2.27 & 2.43 & 2.60 & 2.78 \cr
$\rho$ & 2.20 & 2.36 & 2.52 & 2.70 \cr
$\omega$ & 2.20 & 2.35 & 2.52 & 2.70 \cr
$K^*$(892) & 2.16 & 2.31 & 2.48 & 2.66 \cr
N & 1.57 & 1.69 & 1.81 & 1.95 \cr
$\eta'$(958) & 2.15 & 2.30 & 2.46 & 2.63 \cr
$f_0$(980) & 2.14 & 2.29 & 2.45 & 2.63 \cr
$a_0$(980) & 2.14 & 2.29 & 2.45 & 2.62 \cr
$\phi$(1020) & 2.13 & 2.28 & 2.44 & 2.61 \cr
$\Lambda$ & 1.53 & 1.65 & 1.77 & 1.90 \cr
$h_1$(1170) & 2.10 & 2.24 & 2.40 & 2.57 \cr
$\Sigma$ & 1.52 & 1.63 & 1.75 & 1.88 \cr
$b_1$(1235) & 2.09 & 2.23 & 2.38 & 2.55 \cr
$\Delta$(1232) & 1.51 & 1.62 & 1.74 & 1.87 \cr
$a_1$(1260) & 2.09 & 2.23 & 2.38 & 2.55 \cr
$K_1$(1270) & 2.08 & 2.22 & 2.37 & 2.54 \cr
$f_2$(1270) & 2.08 & 2.22 & 2.37 & 2.54 \cr
$f_1$(1285) & 2.08 & 2.22 & 2.37 & 2.54 \cr
$\eta$(1295) & 2.08 & 2.22 & 2.37 & 2.53 \cr
$\pi$(1300) & 2.08 & 2.22 & 2.37 & 2.53 \cr
$\Xi$ & 1.50 & 1.61 & 1.73 & 1.85 \cr
$a_2$(1320) & 2.08 & 2.21 & 2.36 & 2.53 \cr
$\Sigma$(1385) & 1.49 & 1.60 & 1.71 & 1.84 \cr
$f_0$(1370) & 2.07 & 2.20 & 2.35 & 2.52 \cr
$K_1$(1400) & 2.07 & 2.20 & 2.34 & 2.51 \cr
$\Lambda$(1405) & 1.49 & 1.59 & 1.71 & 1.83 \cr
$K^*$(1410) & 2.06 & 2.20 & 2.34 & 2.51 \cr
$\eta$(1405) & 2.06 & 2.19 & 2.34 & 2.50 \cr
$\omega$(1420) & 2.06 & 2.19 & 2.34 & 2.50 \cr
$f_1$(1420) & 2.06 & 2.19 & 2.34 & 2.50 \cr
$K_0^*$(1430) & 2.06 & 2.19 & 2.34 & 2.50 \cr
$K_2^*$(1430) & 2.06 & 2.19 & 2.34 & 2.50 \cr
N(1440) & 1.49 & 1.59 & 1.70 & 1.83 \cr
$\rho$(1450) & 2.06 & 2.19 & 2.33 & 2.49 \cr
$f_0$(1500) & 2.05 & 2.18 & 2.33 & 2.49 \cr
$\Lambda$(1520) & 1.48 & 1.58 & 1.69 & 1.81 \cr
N(1520) & 1.48 & 1.58 & 1.69 & 1.81 \cr
$f_2'$(1525) & 2.05 & 2.18 & 2.32 & 2.48 \cr
$\Xi$(1530) & 1.48 & 1.58 & 1.69 & 1.81 \cr
N(1535) & 1.47 & 1.58 & 1.69 & 1.81 \cr
$\Delta$(1600) & 1.47 & 1.57 & 1.68 & 1.80 \cr
$\Lambda$(1600) & 1.47 & 1.57 & 1.68 & 1.80 \cr
$\Delta$(1620) & 1.47 & 1.57 & 1.68 & 1.80 \cr
$\omega$(1650) & 2.04 & 2.16 & 2.30 & 2.46 \cr
N(1650) & 1.46 & 1.56 & 1.67 & 1.79 \cr
$\Omega$ & 1.46 & 1.56 & 1.67 & 1.79 \cr
\hline \hline
\end{tabular}
\end{center}
\end{table}

\begin{table}
\caption{Species-dependent shear viscous phase space corrections calculated 
as a function of temperature
for a gas of hadrons up to $m=1.672$~GeV
with the same constant cross section for all species,
assuming linear momentum dependence $\delta f/f^{eq} \propto p$.}
\label{Table:C_linear_constant}
\begin{center}
\begin{tabular}{lcccc}
\hline\hline
Species   & T = 100 & 120  & 140 & 165 MeV\cr
\hline
$\pi$ & 5.81 & 6.06 & 6.29 & 6.49 \cr
K & 5.78 & 6.03 & 6.27 & 6.46 \cr
$\eta$ & 5.79 & 6.03 & 6.27 & 6.46 \cr
$f_0$ & 5.79 & 6.03 & 6.27 & 6.46 \cr
$\rho$ & 5.82 & 6.04 & 6.27 & 6.46 \cr
$\omega$ & 5.82 & 6.04 & 6.27 & 6.46 \cr
$K^*$(892) & 5.85 & 6.06 & 6.28 & 6.46 \cr
N & 5.86 & 6.07 & 6.28 & 6.46 \cr
$\eta'$(958) & 5.87 & 6.07 & 6.28 & 6.46 \cr
$f_0$(980) & 5.87 & 6.07 & 6.29 & 6.46 \cr
$a_0$(980) & 5.87 & 6.07 & 6.29 & 6.46 \cr
$\phi$(1020) & 5.88 & 6.08 & 6.29 & 6.46 \cr
$\Lambda$ & 5.91 & 6.10 & 6.30 & 6.47 \cr
$h_1$(1170) & 5.93 & 6.11 & 6.31 & 6.47 \cr
$\Sigma$ & 5.94 & 6.11 & 6.31 & 6.47 \cr
$b_1$(1235) & 5.95 & 6.12 & 6.32 & 6.48 \cr
$\Delta$(1232) & 5.95 & 6.12 & 6.32 & 6.48 \cr
$a_1$(1260) & 5.95 & 6.12 & 6.32 & 6.48 \cr
$K_1$(1270) & 5.96 & 6.13 & 6.32 & 6.48 \cr
$f_2$(1270) & 5.96 & 6.13 & 6.32 & 6.48 \cr
$f_1$(1285) & 5.97 & 6.14 & 6.32 & 6.48 \cr
$\eta$(1295) & 5.97 & 6.14 & 6.33 & 6.48 \cr
$\pi$(1300) & 5.97 & 6.14 & 6.33 & 6.48 \cr
$\Xi$ & 5.98 & 6.14 & 6.33 & 6.48 \cr
$a_2$(1320) & 5.98 & 6.14 & 6.33 & 6.48 \cr
$\Sigma$(1385) & 6.00 & 6.16 & 6.34 & 6.49 \cr
$f_0$(1370) & 6.00 & 6.16 & 6.34 & 6.49 \cr
$K_1$(1400) & 6.01 & 6.16 & 6.34 & 6.49 \cr
$\Lambda$(1405) & 6.01 & 6.17 & 6.34 & 6.49 \cr
$K^*$(1410) & 6.01 & 6.17 & 6.34 & 6.49 \cr
$\eta$(1405) & 6.01 & 6.17 & 6.34 & 6.49 \cr
$\omega$(1420) & 6.02 & 6.17 & 6.35 & 6.49 \cr
$f_1$(1420) & 6.02 & 6.17 & 6.35 & 6.49 \cr
$K_0^*$(1430) & 6.02 & 6.17 & 6.35 & 6.49 \cr
$K_2^*$(1430) & 6.02 & 6.17 & 6.35 & 6.49 \cr
N(1440) & 6.02 & 6.17 & 6.35 & 6.49 \cr
$\rho$(1450) & 6.03 & 6.18 & 6.35 & 6.50 \cr
$f_0$(1500) & 6.05 & 6.19 & 6.36 & 6.50 \cr
$\Lambda$(1520) & 6.05 & 6.19 & 6.36 & 6.50 \cr
N(1520) & 6.05 & 6.20 & 6.36 & 6.50 \cr
$f_2'$(1525) & 6.05 & 6.20 & 6.36 & 6.50 \cr
$\Xi$(1530) & 6.06 & 6.20 & 6.37 & 6.50 \cr
N(1535) & 6.06 & 6.20 & 6.37 & 6.50 \cr
$\Delta$(1600) & 6.08 & 6.22 & 6.38 & 6.51 \cr
$\Lambda$(1600) & 6.08 & 6.22 & 6.38 & 6.51 \cr
$\Delta$(1620) & 6.09 & 6.22 & 6.38 & 6.51 \cr
$\omega$(1650) & 6.10 & 6.23 & 6.39 & 6.52 \cr
N(1650) & 6.10 & 6.23 & 6.39 & 6.52 \cr
$\Omega$ & 6.11 & 6.24 & 6.39 & 6.52 \cr
\hline \hline
\end{tabular}
\end{center}
\end{table}

\begin{table}
\caption{Species-dependent shear viscous phase space corrections calculated 
as a function of temperature
for a gas of hadrons up to $m=1.672$~GeV
with additive quark model\cite{AQM} (AQM) cross sections,
assuming linear momentum dependence $\delta f/f^{eq} \propto p$.}
\label{Table:C_linear_AQM}
\begin{center}
\begin{tabular}{lcccc}
\hline\hline
Species & T = 100 & 120  & 140 & 165 MeV\cr
\hline
$\pi$ & 5.84 & 6.15 & 6.49 & 6.84 \cr
K & 5.81 & 6.12 & 6.46 & 6.81 \cr
$\eta$ & 5.81 & 6.12 & 6.46 & 6.81 \cr
$f_0$ & 5.82 & 6.12 & 6.46 & 6.81 \cr
$\rho$ & 5.85 & 6.13 & 6.46 & 6.81 \cr
$\omega$ & 5.85 & 6.13 & 6.46 & 6.81 \cr
$K^*$(892) & 5.88 & 6.15 & 6.47 & 6.81 \cr
N & 4.25 & 4.47 & 4.72 & 4.97 \cr
$\eta'$(958) & 5.89 & 6.16 & 6.48 & 6.81 \cr
$f_0$(980) & 5.90 & 6.16 & 6.48 & 6.81 \cr
$a_0$(980) & 5.90 & 6.16 & 6.48 & 6.81 \cr
$\phi$(1020) & 5.91 & 6.17 & 6.48 & 6.82 \cr
$\Lambda$ & 4.27 & 4.47 & 4.72 & 4.96 \cr
$h_1$(1170) & 5.96 & 6.20 & 6.50 & 6.82 \cr
$\Sigma$ & 4.28 & 4.48 & 4.72 & 4.96 \cr
$b_1$(1235) & 5.98 & 6.21 & 6.51 & 6.83 \cr
$\Delta$(1232) & 4.29 & 4.48 & 4.72 & 4.96 \cr
$a_1$(1260) & 5.98 & 6.21 & 6.51 & 6.83 \cr
$K_1$(1270) & 5.99 & 6.22 & 6.52 & 6.83 \cr
$f_2$(1270) & 5.99 & 6.22 & 6.52 & 6.83 \cr
$f_1$(1285) & 5.99 & 6.23 & 6.52 & 6.83 \cr
$\eta$(1295) & 6.00 & 6.23 & 6.52 & 6.83 \cr
$\pi$(1300) & 6.00 & 6.23 & 6.52 & 6.83 \cr
$\Xi$ & 4.30 & 4.49 & 4.72 & 4.96 \cr
$a_2$(1320) & 6.01 & 6.23 & 6.52 & 6.84 \cr
$\Sigma$(1385) & 4.31 & 4.50 & 4.72 & 4.96 \cr
$f_0$(1370) & 6.02 & 6.25 & 6.53 & 6.84 \cr
$K_1$(1400) & 6.04 & 6.25 & 6.54 & 6.84 \cr
$\Lambda$(1405) & 4.31 & 4.50 & 4.73 & 4.96 \cr
$K^*$(1410) & 6.04 & 6.26 & 6.54 & 6.85 \cr
$\eta$(1405) & 6.04 & 6.26 & 6.54 & 6.85 \cr
$\omega$(1420) & 6.04 & 6.26 & 6.54 & 6.85 \cr
$f_1$(1420) & 6.05 & 6.26 & 6.54 & 6.85 \cr
$K_0^*$(1430) & 6.05 & 6.26 & 6.54 & 6.85 \cr
$K_2^*$(1430) & 6.04 & 6.26 & 6.54 & 6.85 \cr
N(1440) & 4.32 & 4.50 & 4.73 & 4.96 \cr
$\rho$(1450) & 6.06 & 6.27 & 6.55 & 6.85 \cr
$f_0$(1500) & 6.07 & 6.28 & 6.56 & 6.86 \cr
$\Lambda$(1520) & 4.33 & 4.51 & 4.73 & 4.97 \cr
N(1520) & 4.33 & 4.51 & 4.73 & 4.97 \cr
$f_2'$(1525) & 6.08 & 6.29 & 6.56 & 6.86 \cr
$\Xi$(1530) & 4.34 & 4.51 & 4.73 & 4.97 \cr
N(1535) & 4.34 & 4.51 & 4.73 & 4.97 \cr
$\Delta$(1600) & 4.35 & 4.52 & 4.74 & 4.97 \cr
$\Lambda$(1600) & 4.35 & 4.52 & 4.74 & 4.97 \cr
$\Delta$(1620) & 4.35 & 4.52 & 4.74 & 4.97 \cr
$\omega$(1650) & 6.13 & 6.32 & 6.59 & 6.87 \cr
N(1650) & 4.36 & 4.53 & 4.74 & 4.97 \cr
$\Omega$ & 4.36 & 4.53 & 4.74 & 4.97 \cr
\hline \hline
\end{tabular}
\end{center}
\end{table}

\begin{table}
\caption{Parameters as a function of temperature for the fit function
 $c(x) = \delta + \alpha / [1 + (x/\gamma)^\beta]$ to 
the species-dependent Grad shear viscous phase space 
corrections listed in Tables~\ref{Table:C_Grad_constant}-\ref{Table:C_Grad_AQM}.}
\label{Table:Cfits_Grad}
\begin{center}
\begin{tabular}{ccccc}
\hline\hline \\[-0.2cm]
\multicolumn{5}{c}{ $\sigma = const$ scenario (Grad)} \\[0.1cm]
$T$~[MeV] & $\alpha$ & $\beta$ & $\gamma$ & $\delta$ \cr
\hline
       100 &  0.698 &  1.204 &  0.715 &  0.467 \cr
       120 &  0.700 &  1.266 &  0.862 &  0.493 \cr
       140 &  0.702 &  1.326 &  0.996 &  0.519 \cr
       165 &  0.693 &  1.397 &  1.140 &  1.243 \cr
\hline \\[-0.2cm]
\multicolumn{5}{c}{ AQM scenario, mesons (Grad)} \\[0.1cm]
$T$~[MeV] & $\alpha$ & $\beta$ & $\gamma$ & $\delta$ \cr
\hline
       100 &  0.696 &  1.214 &  0.712 &  0.472 \cr
       120 &  0.704 &  1.278 &  0.856 &  0.505 \cr
       140 &  0.715 &  1.342 &  0.985 &  0.543 \cr
       165 &  0.717 &  1.414 &  1.124 &  0.591 \cr
\hline \\[-0.2cm]
\multicolumn{5}{c}{ AQM scenario, baryons (Grad)} \\[0.1cm]
$T$~[MeV] & $\alpha$ & $\beta$ & $\gamma$ & $\delta$ \cr
\hline
       100 &  0.687 &  1.009 &  0.623 &  0.286 \cr
       120 &  0.698 &  1.037 &  0.801 &  0.297 \cr
       140 &  0.710 &  1.075 &  0.987 &  0.311 \cr
       165 &  0.711 &  1.129 &  1.204 &  0.334 \cr
\hline
\hline
\end{tabular}
\end{center}
\end{table}

\begin{table}
\caption{Parameters as a function of temperature for the fit function
 $c(x) = \delta + \alpha / [1 + (x/\gamma)^\beta]$ to 
the species-dependent $p^{3/2}$ shear viscous phase space 
corrections listed in 
Tables~\ref{Table:C_3over2_constant}-\ref{Table:C_3over2_AQM}.}
\label{Table:Cfits_3over2}
\begin{center}
\begin{tabular}{ccccc}
\hline\hline \\[-0.2cm]
\multicolumn{5}{c}{ $\sigma = const$ scenario ($p^{3/2}$)} \\[0.1cm]
$T$~[MeV] & $\alpha$ & $\beta$ & $\gamma$ & $\delta$ \cr
\hline
       100 &  0.748 &  1.446 &  0.559 &  1.900 \cr
       120 &  0.823 &  1.375 &  0.712 &  1.933 \cr
       140 &  0.883 &  1.363 &  0.876 &  1.969 \cr
       165 &  0.921 &  1.386 &  1.061 &  2.006 \cr
\hline \\[-0.2cm]
\multicolumn{5}{c}{ AQM scenario, mesons ($p^{3/2}$)} \\[0.1cm]
$T$~[MeV] & $\alpha$ & $\beta$ & $\gamma$ & $\delta$ \cr
\hline
       100 &  0.759 &  1.427 &  0.561 &  1.904 \cr
       120 &  0.839 &  1.370 &  0.714 &  1.959 \cr
       140 &  0.908 &  1.367 &  0.874 &  2.032 \cr
       165 &  0.957 &  1.397 &  1.047 &  2.126 \cr
\hline \\[-0.2cm]
\multicolumn{5}{c}{ AQM scenario, baryons ($p^{3/2}$)} \\[0.1cm]
$T$~[MeV] & $\alpha$ & $\beta$ & $\gamma$ & $\delta$ \cr
\hline
       100 &  0.540 &  1.627 &  0.760 &  1.344 \cr
       120 &  0.691 &  1.396 &  0.836 &  1.369 \cr
       140 &  0.806 &  1.288 &  0.974 &  1.400 \cr
       165 &  0.890 &  1.243 &  1.177 &  1.439 \cr
\hline \hline
\end{tabular}
\end{center}
\end{table}

\begin{table}
\caption{Parameters as a function of temperature for the fit function
 $c(x) = \alpha + \beta | x - \gamma|^\delta$ to 
the species-dependent $p^1$ shear viscous phase space 
corrections listed in 
Tables~\ref{Table:C_linear_constant}-\ref{Table:C_linear_AQM}.}
\label{Table:Cfits_linear}
\begin{center}
\begin{tabular}{ccccc}
\hline\hline \\[-0.2cm]
\multicolumn{5}{c}{ $\sigma = const$ scenario ($p^1$)} \\[0.1cm]
$T$~[MeV] & $\alpha$ & $\beta$ & $\gamma$ & $\delta$ \cr
\hline
       100 &  5.775 &  0.240 &  0.419 &  1.521 \cr
       120 &  6.025 &  0.166 &  0.502 &  1.633 \cr
       140 &  6.265 &  0.114 &  0.599 &  1.734 \cr
       165 &  6.458 &  0.073 &  0.747 &  1.882 \cr
\hline \\[-0.2cm]
\multicolumn{5}{c}{ AQM scenario, mesons ($p^1$)} \\[0.1cm]
$T$~[MeV] & $\alpha$ & $\beta$ & $\gamma$ & $\delta$ \cr
\hline
       100 &  5.802 &  0.239 &  0.419 &  1.546 \cr
       120 &  6.114 &  0.167 &  0.504 &  1.655 \cr
       140 &  6.459 &  0.118 &  0.603 &  1.743 \cr
       165 &  6.808 &  0.080 &  0.742 &  1.865 \cr
\hline \\[-0.2cm]
\multicolumn{5}{c}{ AQM scenario, baryons ($p^1$)} \\[0.1cm]
$T$~[MeV] & $\alpha$ & $\beta$ & $\gamma$ & $\delta$ \cr
\hline
       100 &  4.245 &  0.156 &  0.848 &  1.404 \cr
       120 &  4.466 &  0.097 &  0.897 &  1.603 \cr
       140 &  4.715 &  0.062 &  1.015 &  1.842 \cr
       165 &  4.963 &  0.045 &  1.293 &  1.931 \cr
\hline \hline
\end{tabular}
\end{center}
\end{table}


\end{document}